\documentstyle[aps,prb,epsf,psfig]{revtex}

\begin{document}

\title{Phase diffusion and suppression of the supercurrent by  
  quantum-mechanical fluctuations of the Josephson plasma at 
  small superconducting point contacts}
        
\author{Kurt Gloos $^{1,2}$ and Frithjof Anders $^1$}
\address{$^1$ Institut f\"ur Festk\"orperphysik,
 Technische Universit\"at Darmstadt, D-64289 Darmstadt, Germany}
\address{$^2$ Max-Planck-Institut f\"ur chemische Physik fester Stoffe,
 D-01187 Dresden, Germany}

\date{March 8, 1999}

\maketitle

\begin{abstract}
The RCSJ model of resistively and capacitively shunted Josephson junctions 
is used to describe superconducting point contacts over a wide range of 
resistances up to the metallic -- tunneling transition. Their small dynamic 
capacitance of order $C = 0.1\,$fF due to the point-contact geometry results 
in a huge plasma frequency. The critical current is then strongly suppressed 
and the contact resistance becomes finite because of quantum-mechanical 
zero-point fluctuations of the Josephson plasma and the rather large escape 
rate out of the zero-voltage state due to quantum tunneling. We test the
predictions of the RCSJ model on the classical superconductors lead, indium, 
aluminum, and cadmium.
\end{abstract}




\section{Overview}

Point contacts between two superconductors have attracted interest for a 
number of reasons: They can be used to study the behaviour of a macroscopic 
quantum object \cite{Martinis87} as well as quantum transport when the 
junction consists of few atoms only \cite{Cuevas96,LevyYeyati96}.
They may also help to
extract information on the order parameter of compounds like the 
heavy-fermion superconductors, although those experiments are not as easy to 
interpret as has previously been assumed \cite{Gloos-latsize}.

Metallic point contacts of classical superconductors have first been 
investigated systematically by Muller {\em et al. }
\cite{Muller92,Muller94,Peters95,vanderPost97}. They observed that, on
reducing the contact area, the residual contact resistance $R_0$ at zero 
bias becomes finite and the critical current $I_{\rm{c}}$ smaller than 
the theoretical value $I_{\rm{c}}^0$ at a temperature $T \ll T_{\rm{c}}$
even when the normal-state resistance $R_{\rm{N}} = R(T>T_{\rm{c}})$ is 
far below the quantum limit $R_{\rm{K}}/2=h/2e^2=12.9\,\rm{k}\Omega$. 
This was attributed to the presence of  electrical radio-frequency 
noise and additional damping of the junctions at high frequencies due to 
the small impedance (that is the large capacitance) of the current and 
voltage leads \cite{Peters95,vanderPost97}. The Andreev-reflection 
excess current, on the other hand, was found not to depend on the 
lateral size of the junctions. 

We show here that the behaviour of those Josephson junctions can better be
described by the small capacitance $C$ typical for the 
three-dimensional point-contact 
geometry. Taking into account the capacitance  of the junctions as an 
adjustable parameter, we derive relations between $R_{\rm{N}}$ and the 
residual resistance $R_0$ as well as the product $R_{\rm{N}} I_{\rm{c}}$, 
and estimate both the plasma frequency $\omega_{\rm{p}}$ and the 
Josephson coupling energy $E_{\rm{JE}}$ from the current-voltage $I(U)$ 
characteristic at $I \rightarrow 0$. Our experiments on the classical 
superconductors lead, indium, aluminum, and cadmium fit the predictions 
fairly well. Using the horizon model, the capacitance $C$ can be
traced back to the properties of vacuum tunneling junctions in the 
normal state. 

\section{Theory}

The system of Cooper pairs in a bulk sample forms one wavefunction 
$\Psi = |\Psi_0| \exp{(i \varphi)}$. The gradient of the phase $\varphi$ 
drives the current. According to B.~D.~Josephson \cite{Josephson62}, by 
coupling two superconductors weakly to each other using a thin insulating 
layer between them, the phase difference $\varphi := \varphi_2 - \varphi_1$ 
across the junction results in a supercurrent 
\begin{equation}
I = I_{\rm{c}}^0\sin{\varphi}
\end{equation}
as long as the voltage drop $U$ is much smaller than the superconducting 
energy gap $2\Delta (T)/e$. According to Ambegaokar and Baratoff 
\cite{Ambegaokar63} the critical current of a Josephson tunnel junction
$I_{\rm{c}}^0 = (\pi\Delta/2eR_{\rm{N}})\tanh{\left(\Delta/2k_{\rm{B}}
T\right)}$.

The critical current through short 
metallic junctions between isotropic BCS-type superconductors was derived by 
Kulik and Omel'yanchuk to be (KO1) $I_{\rm{c}}^0 = (1.32\,\pi\Delta/
2eR_{\rm{N}}) \tanh{\left(\Delta/2k_{\rm{B}} T\right)}$ \cite{KO1} in the 
dirty limit (electron mean free path $l$ much shorter than the 
superconducting coherence length $\xi$), and (KO2) $I_{\rm{c}}^0 = 
(\pi\Delta/ e R_{\rm{N}}) \tanh{\left(\Delta/2k_{\rm{B}} T\right)}$
\cite{KO2} in the clean limit  $l \gg \xi$. We will see below that quantum
fluctuations of the Josephson plasma can further reduce the critical current.
The superscript 0 denotes then the above intrinsic value.

The current-phase relationship of a metallic Josephson junction in the
clean limit deviates from the simple sinusoidal of a tunnel junction.
It depends on the  transmission coefficient $D$ as \cite{Haberkorn78}
\begin{equation}
I(\varphi) = \frac {\pi\Delta} {2eR_{\rm{N}}} 
            \frac {\sin{\varphi}} {\delta} 
            \tanh{\left( \frac {\Delta\delta} {2k_{\rm{B}} T } \right)}
\label{cpr}
\end{equation}
with $\delta = \sqrt{1-D \sin{^2 \left(\varphi/2\right)}}$. The 
normal-state contact resistance in the ballistic limit
\begin{equation}
R_{\rm{N}} = \frac{2R_{\rm{K}}}{(ak_{\rm{F}})^2} D^{-1}
\label{ballistic}
\end{equation}
Here $k_{\rm{F}}$ is the Fermi wave number and $a$ the contact radius.
By varying the transmission coefficient, the junction can be tuned
from pure metallic ($D = 1$) to pure tunnel ($D = 0$) behaviour.

Note  the effects we are going to investigate do not involve an 'energy 
cost' due to lateral confinement that could suppress the superconducting 
features in the contact region: although the junctions are rather 
small -- with a typical metallic $k_{\rm{F}} \approx 15\,{\rm{nm}}^{-1}$
a $R_{\rm{N}}=100\,\Omega$ junction has a contact radius $a$ of about
2 nm -- the Heisenberg uncertainty principle yields a spread with respect
to momentum, and not one with respect to energy.

\begin{figure}[tb]
\hspace{20mm}\psfig{file=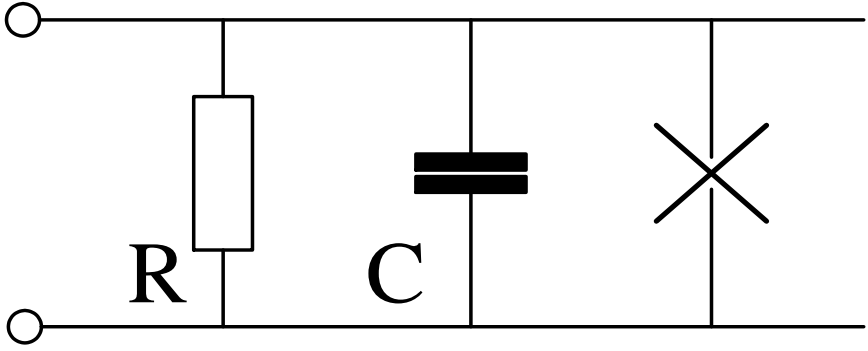,angle=270,width=120mm}
\caption[RCSJ model-x]{\small \rm 
RCSJ model. The junction itself is represented by $\times$. The current is
applied to and the voltage detected at the terminals denoted by $\circ$.}
\label{RCSJ model}
\end{figure}

\subsection{The RCSJ model and the quasi-particle resistance}
When the applied voltage $U = 0$, the phase does not change because 
$\dot{\varphi} = \partial \varphi / \partial t =  2 e U / \hbar = 0$,
and the current through the junction is constant \cite{Josephson62}.
If the voltage is finite, the phase increases at a rate 
$2 e U /\hbar \approx 3\cdot 10^{15}\,(1/{\rm{s}})\cdot U({\rm{V}})$, 
and the current has an oscillatory component at extremely high frequencies. 
D.~E.~McCumber \cite{McCumber68} and W.~C.~Stewart \cite{Stewart68} took 
into account that besides the supercurrent across the Josephson tunnel
junction there exist also a normal quasi-particle current, described by a
parallel resistance $R=R_{\rm{qp}}$,
as well as a displacement current due to a (stray) capacitance $C$
(Fig.~\ref{RCSJ model}). This capacitance is important for the properties 
of the junction because it represents a short circuit for the high-frequency 
part of the supercurrent. The total current of such a resistively and 
capacitively shunted junction (RCSJ) is 
$I = I_{\rm{c}}^0\sin{\varphi} + U/R_{\rm{qp}} + CdU/dt$ or 
\begin{equation}
\frac{I}{I_{\rm{c}}^0}=\sin{\varphi} + \frac{\dot{\varphi}}{\omega_{\rm{c}}}
  + \frac{\ddot{\varphi}}{\omega_{\rm{p}}^2}
\label{jo-rcsj}
\end{equation}
with the plasma frequency        
$\omega_{\rm{p}} = \sqrt{2 e I_{\rm{c}}^0 /\hbar C}$
and the characteristic frequency 
$\omega_{\rm{c}} = 2 e I_{\rm{c}}^0 R_{\rm{qp}} / \hbar$.

The quasi-particle resistance $R=R_{\rm{qp}}$ is {\em a priori} unknown.
Generally, it will depend both on voltage, on current, or on phase. 
Above $T_{\rm{c}}$, without Josephson effect, the quasi-particle resistance
equals just the normal-state contact resistance $R_{\rm{N}}$, 
originating purely from the quasi-particle current through the junction 
(we assume there is no external shunt resistance). For a tunnel
junction, this $R_{\rm{qp}}$ is proportional to the square modulus of
the tunnel matrix element. For a ballistic junction, like the ones we
are using, $R_{\rm{qp}}$ involves an infinite resummation of multiple
quasi-particle transmissions through the contact. Then, energy is not
dissipated at the junction itself: electrons equilibriate far away
from the junction in the bulk metal. By including non-linear effects
like quasi-particle quasi-particle interactions or electron-phonon
coupling, the contact  resistance, and thus $R_{\rm{qp}}$, depends on
temperature and the applied bias voltage $U$.

Below $T_{\rm{c}}$, the situation changes completely as a Josephson 
supercurrent can flow through the junction. However, well below $T_{\rm{c}}$
there is no quasi-particle density of states within the energy gap
$2\Delta$ on both sides of the junction.
Therefore, $R_{\rm{qp}} \rightarrow \infty$ for an ideal
junction at low bias voltage $|U| \ll 
\Delta/e$. At a Josephson tunnel  junction, $R_{\rm{qp}}$  can be
directly observed by suppressing the supercurrent using a small
magnetic field less than the critical field. Tunnel experiments have
shown that the damping resistance can be much maller  than the
intrinsic quasi-particle resistance $R_{\rm{qp}}$, see for example
Ref.~\cite{Martinis87}. This is probably due to parasitic damping of
the high-frequency electromagnetic field generated at the junction.  

The quasi-particle resistance of a metallic Josephson junction (or
point contact) can not be measured directly, as far as we know. But it
seems plausible, to apply the same arguing as for the tunnel
junctions: For a clean BCS-type superconductor far below $T_{\rm{c}}$,
the quasi-particle density of states has vanished in the vicinity of
the chemical potential, and therefore the
intrinsic $R_{\rm{qp}} \rightarrow \infty$. Recently, Levy Yeyati {\em
et al.~}\cite{LevyYeyati96} derived the dissipative part of the
current through a superconducting quantum point contact, that is a
ballistic junction with one conducting channel. Assuming the same
relation to hold also in the general case of a ballistic Josephson
junction, the invers quasi-particle resistance
\begin{equation}                                           
    R_{\rm{qp}}^{-1}(\varphi) = \frac {D} {R_{\rm{N}}}
    \frac {\pi\Delta^2} {16\Gamma k_{\rm{B}}T}
    \left[
    \frac {\sin{\varphi}} {\delta}\,
    {\rm{sech}} \left( \frac {\Delta\delta}{2k_{\rm{B}}T} \right) 
    \right]^2
\label{qp-damping}
\end{equation}
$\Gamma = \hbar/\tau$ being the inelastic relaxation rate, and $\tau$ the
quasi-particle lifetime. Most importantly, the dissipative part of the current
vanishes at $\varphi \rightarrow 0$ or at $I \ll I_{\rm{c}}$, that is
$R_{\rm{qp}} \gg R_{\rm{N}}$.

In both type of experiments, Josephson tunnel junctions as well as 
metallic Josephson point contacts, there can be damping due to the external 
circuitry. The amount of this parasitic 
damping can not be predicted theoretically, because it depends on the 
specific experimental setup. Thus the damping resistance
$R = R_{\rm{qp}}$ of the RCSJ model has to be treated as a free parameter,
like the capacitance, that has to be extracted from the properties of the
junctions. The normal-state contact resistance may represent some lower bound
of the intrinsic quasi-particle resistance, and $R_{\rm{N}}$ may thus serve
as a first guess for the damping resistance. But otherwise there is no reason to set $R_{\rm{qp}} = R_{\rm{N}}$.

\begin{figure}
\hspace{15mm}\psfig{file=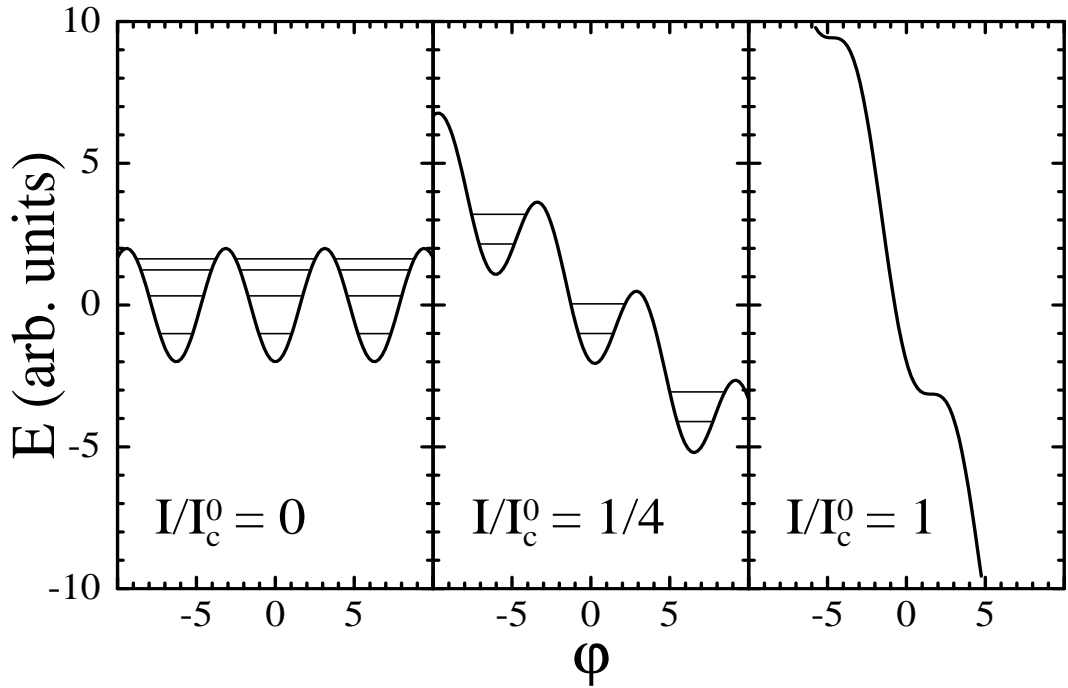,angle=270,width=150mm}
\caption[washboard-x]{\small \rm 
Washboard potential $E(I,\varphi)$ of Eq.~\ref{jo-washboard} for different 
values of the applied current.}
\label{washboard}
\end{figure}

Eq.~\ref{jo-rcsj} describes the motion of a particle with 'mass'
$(\hbar/2e)^2 C$ exposed to a viscous damping force 
$(\hbar/2 e R_{\rm{qp}})\dot{\varphi}$ in the potential 
\begin{equation}
E(I,\varphi)= -\frac{\hbar}{2 e } I\varphi - 
    E_{{\rm{JE}}}\cdot \cos{\varphi}
\label{jo-washboard}
\end{equation}
$E_{\rm{JE}}=\hbar I_{\rm{c}}^0/2 e$ is called Josephson coupling energy.
The potential Eq.~\ref{jo-washboard} has the shape of a washboard, its wells 
have a depth of $2E_{\rm{JE}}$ at $I=0$ (Fig.~\ref{washboard}). The number
of discrete energy levels in these wells amounts to about \cite{Martinis87} 
${\rm{Int}}(2E_{\rm{JE}}/\hbar\omega_{\rm{p}})$. The lowest one is the 
zero-point energy $\epsilon_0=\hbar\omega_{\rm{p}}/2$. Minima exist
as long as
$I<I_{\rm{c}}^0$, and the static solution of Eq.~\ref{jo-rcsj} is
$I/I_{\rm{c}}^0=\sin{\varphi}$. If the time constant $R_{\rm{qp}}C$ is much 
smaller than the period of the plasma oscillations, the system is 
overdamped. The only stable state is then at such a potential minimum and 
the $I(U)$ - characteristic is non-hysteretic. An underdamped junction can 
either be at a minimum, the supercurrent flowing without a voltage drop. 
Or it can run down the washboard, and the current is accompanied by a 
voltage. The $I(U)$ - characteristic is then hysteretic. At $I>I_{\rm{c}}^0$ 
there are no more minima, and the system is definitely in the resistive 
state with $\dot{\varphi}\not= 0$. 

The RCSJ model  has been applied to weakly damped planar tunnel junctions 
(with an insulating oxide layer instead of a vacuum gap) of classical 
superconductors like niobium and tin: Fulton and Dunkleberger 
\cite{Fulton74} observed thermally activated escape from the minima, Voss 
and Webb \cite{Voss81} and later on Washburn {\em et al.} \cite{Washburn85} 
reported macroscopic quantum tunneling through the wells of the washboard 
potential. Its discrete energy levels have been observed by Martinis 
{\em et al.} \cite{Martinis85}. Obviously, the RCSJ model is the minimum 
system to study the Josephson effect at solitary point contacts.

\begin{figure}[t]
\psfig{file=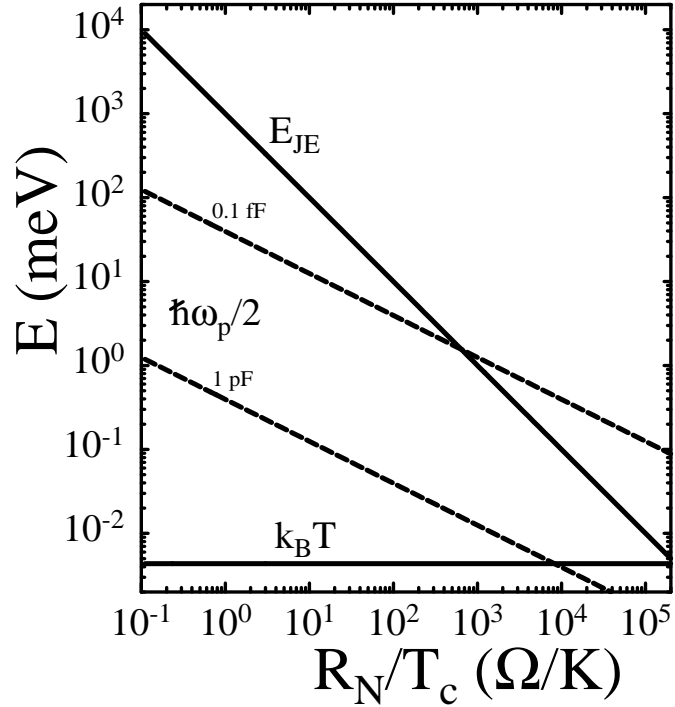,angle=270,width=150mm}
\caption[energies-x]{\small \rm 
Josephson coupling energy $E_{\rm{JE}}$, zero-point energy 
$\hbar\omega_{\rm{p}}/2$ at two different capacitances $C$, and thermal 
energy $k_{\rm{B}}T$ at 50 mK vs.~$R_{\rm{N}}/T_{\rm{c}}$ of junctions
between BCS-type superconductors in the clean limit.}
\label{energies}
\end{figure}

Let us illustrate the situation by setting in numbers. For junctions with 
BCS-type superconductors in the clean limit the coupling energy amounts to
\begin{equation}
E_{\rm{JE}} \approx 1\,{\rm{eV}}\, \frac{T_{\rm{c}}({\rm{K}})}
   {R_{\rm{N}}(\Omega)}
\end{equation}
at a zero-point energy of the Josephson plasma
\begin{equation}
\frac{\hbar\omega_{\rm{p}}}{2} \approx 12\,{\rm{meV}}\,
  \sqrt{\frac{T_{\rm{c}}({\rm{K}})}{R_{\rm{N}}(\Omega)C({\rm{fF}})}}
\end{equation}
Planar tunnel junctions typically have capacitances of order 1 pF. For them 
$\hbar\omega_{\rm{p}}/2$ can be orders of magnitude smaller than 
$E_{\rm{JE}}$. To induce 
a detectable escape from the minima of the potential wells, these junctions 
have to be driven near the critical current. Metallic point contacts, on the 
other hand, have considerably smaller capacitances.  They are believed to be 
of order 1 fF \cite{vanderPost97}. And although the coupling energy 
can be much larger than the thermal energy $k_{\rm{B}}T$ (about
$4\,\mu{\rm{eV}}$ at our lowest temperature of $\sim 50\,$ mK), 
the zero-point energy can be quite high, see Fig.~\ref{energies}. This 
has direct consequences with respect to the critical current and the 
residual resistance, because in such a system the particle can escape 
easily from the minima even at small currents, accompanied by $2\pi$ phase
slips. In addition, because of the quality factor 
\begin{equation}
Q = \omega_{\rm{p}}R_{\rm{qp}} C
 \approx 0.04\,R_{\rm{qp}}(\Omega)
 \sqrt{\frac{T_{\rm{c}}({\rm{K}})C({\rm{fF}})} {R_{\rm{N}}(\Omega) }}
\end{equation}
those metallic junctions have to be expected to be in the cross-over region
between weak and strong damping, if the quasi-particle resistance
$R_{\rm{qp}}$ was of the same size as the normal-state resistance
$R_{\rm{N}}$.

\subsection{Critical current suppressed by quantum fluctuations}
The quantum-mechanical treatment of the particle in the washboard
potential describes the phase as an operator conjugated to the charge $q=2e$
of the Cooper pairs, $\left[\varphi,q\right] = 2i$e.
This allows to include quantum fluctuations and phase diffusion. The
macroscopic voltage is then given by the time derivative
of the expectation value of the phase $U=(\hbar/2e)
\partial \left< \varphi (t)\right> /\partial t$. It was shown, for 
example, that Josephson junctions at $T = 0$ can undergo a phase transition 
from the superconducting to the resistive state, depending on the
strength of dissipation or damping \cite{Schmid83}.

A current $I=x I_{\rm{c}}^0$ tilts the washboard potential
(Fig.~\ref{washboard}) and reduces both the minimum height of the wells
\cite{Fulton74}
\begin{equation}
  2E(x) =  E_{\rm{JE}} \left[ x (2\arcsin{x} - \pi) + 2\cos{\arcsin{x}}
                       \right]
\end{equation}
and the plasma frequency 
\begin{equation}
  \omega (x) = \omega_{\rm{p}} \left( 1-x^2 \right) ^{1/4}. 
\label{plasma}
\end{equation}
$I_{\rm{c}}$ can closely approximate $I_{\rm{c}}^0$ at small $R_{\rm{N}}$ 
when $\hbar \omega_{\rm{p}} \ll E_{\rm{JE}}$. The $I(U)$-characteristic shows
then a sudden rise at $I\approx I_{\rm{c}}^0$. At larger resistances, 
$E_{\rm{JE}}$ decreases faster than $\hbar\omega_{\rm{p}}$, and  the 
ratio $\hbar \omega(x) / E(x)$ diverges as $x\rightarrow 1$, that is at
$I \rightarrow I_{\rm{c}}^0$. Assuming the 
supercurrent being suppressed by the quantum-mechanical fluctuations of the 
Josephson plasma, the actual critical current $I_{\rm{c}}$ of the RCSJ model 
less than $I_{\rm{c}}^0$ is reached at
\begin{equation}
2E(x_{\rm{c}}) \approx \hbar\omega(x_{\rm{c}})/2
\label{eje-wp}
\end{equation}
when the lowest energy level coincides with the minimum height of the 
potential well. The phase difference across the junction is then not 
well-defined any more. Eq.~\ref{eje-wp} makes the reduced critical current
$x_{\rm{c}}=I_{\rm{c}}/I_{\rm{c}}^0$ an implicit function of the 
intrinsic $I_{\rm{c}}^0$ and of the capacitance via
\begin{equation}
  I_{\rm{c}}^0 \approx \frac {\pi e} {R_{\rm{K}} C}
  \left[ \frac {E_{\rm{JE}}\omega(x_{\rm{c}})} 
  {\omega_{\rm{p}} E(x_{\rm{c}})} \right] ^2   \;\; .
\label{reduced-ic}
\end{equation}
Because of the quantum fluctuations the $I(U)$ - characteristic is washed out
and also the critical current becomes less well defined. 

Eq.~\ref{reduced-ic} relates the normal-state resistance of a junction to 
its reduced critical current $x_{\rm{c}}$ as 
\begin{equation}
  R_{\rm{N}}(x_{\rm{c}}) \approx  
  R_{\rm{K}} (R_{\rm{N}} I_{\rm{c}}^0) \frac {C}{\pi e}
  \left[ \frac {\omega_{\rm{p}} E(x_{\rm{c}})}{E_{\rm{JE}} 
  \omega(x_{\rm{c}})} \right] ^2 
\label{icrn}
\end{equation}
depending only on the capacitance $C$ and the material-dependent
parameter $(R_{\rm{N}}I_{\rm{c}}^0)$. The supercurrent vanishes completely 
above 
\begin{equation}
  R_{\rm{N}}(x_{\rm{c}}=0) 
  \approx R_{\rm{K}} (R_{\rm{N}} I_{\rm{c}}^0) \frac{C}{\pi e}
\;\; .
\label{critical resistance}
\end{equation} 
Although we have neglected damping, Eq.~\ref{critical resistance}
resembles the $s = \sqrt{4 R_{\rm{K}} C I_{\rm{c}}^{0}/e\pi} = 1.5$ and 
$\eta = R_{\rm{K}}/8\pi R_{\rm{qp}} = 1/(2\pi)$ fixpoint for the 
localization -- delocalization transition at $T\rightarrow 0$, the 
so-called Schmid transition \cite{Schmid83}. It demonstrates that our 
simple physical argument captures the essence of a renomalisation group 
calculation.

\begin{figure}
\hspace{15mm}\psfig{file=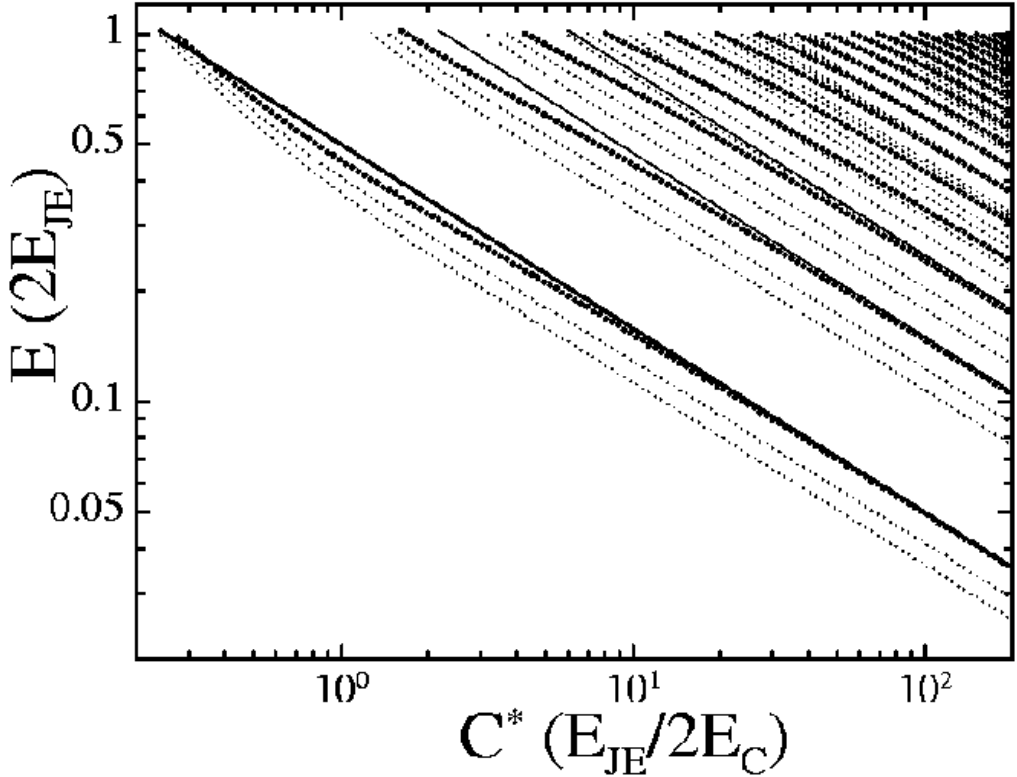,width=150mm}
\caption[energy levels-x]{\small \rm 
Energy of the stationary states in units of $2E_{\rm{JE}}$ vs.~ reduced
mass $C^* = E_{\rm{JE}}/2E_{\rm{C}}$ for tunnel junctions ($D=0$) and
metallic junctions ($D=1$) at low ($T/T_{\rm{c}}=0.01$) and high
($T/T_{\rm{c}}=0.1$) temperatures (from top to bottom). The analytical
zero-point energy $\epsilon_0=2E_{\rm{JE}}\sqrt{E_{\rm{C}}/2E_{\rm{JE}}}$
as well as the first and second excited level (solid lines) fit well the
numerical data for the tunnel junctions ($E_{\rm{C}} =e^2/2C$).}
\label{energy levels}
\end{figure}

A metallic Josephson junction with finite transmission coefficient slightly
changes the situation with respect to a tunnel junction. Neglecting
dissipation, the Schr\"odinger equation
\begin{equation}
  \frac{\partial^2}{\partial\varphi^2}\Psi = \frac{C}{2e^2}
  \left[E(\varphi)-\epsilon\right] \Psi
\end{equation}
can be solved numerically. Fig.~\ref{energy levels} shows the Eigen energy
$\epsilon=\epsilon_i$ of the stationary states $i=0,1,2,...$ fit quite well
the analytical data of the washboard potential Eq.~\ref{jo-washboard} at
small energies. For metallic junctions with the current-phase relationship
of Eq.~\ref{cpr}, the potential minima broaden and the levels shift towards
smaller energies. Since these are small corrections, for
ease of use, and because we do not know the transmission coefficient
of the individual junctions, our discussion will be based on the washboard
potential.

For $R_{\text{N}}I_{\rm{c}}^0 \approx 0.1\,$meV and $C\approx 1\,$fF the
critical resistance of Eq.~\ref{critical resistance} is of order 5 k$\Omega$, 
which is still in the metallic regime. It shows that these phenomena due 
to the capacitance of the junctions have to be taken into account seriously.

\subsection{Phase diffusion: Finite contact resistance and suppression
              of the critical current}
At sufficiently high temperatures thermal activation induces escape from the 
potential minima by $2\pi$ phase slips, that is the phase diffuses. In this
regime, the finite contact resistance due to a small capacitance has already
been discussed by Ambegaokar and Halperin \cite{Ambegaokar69}.

\begin{figure}
\hspace{15mm} \psfig{file=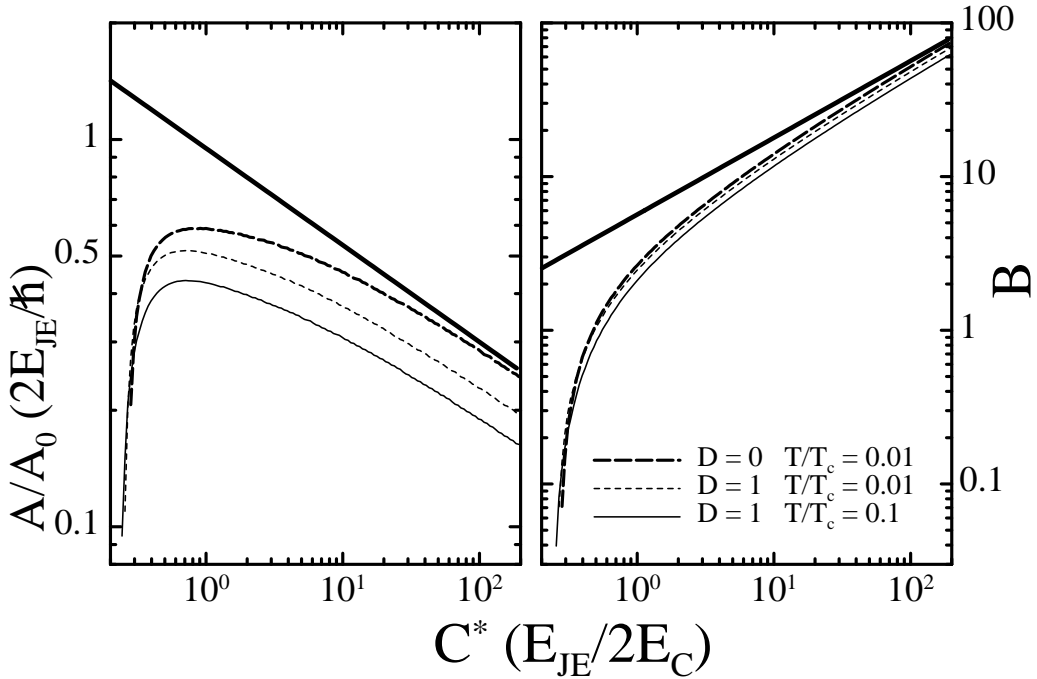,angle=270,width=140mm}
\caption[ab-coefficients-x]{\small \rm 
a) $A/A_0$ and b) $B$ vs.~reduced mass $C^* = E_{\rm{JE}}/2E_{\rm{C}}$ 
for tunnel junctions ($D=0$) and metallic junctions ($D=1$) at 
low ($T/T_{\rm{c}}=0.01$) and high ($T/T_{\rm{c}}=0.1$) temperatures. 
The analytical $A/A_0=2 \epsilon_0 \sqrt{B/2\pi}/\hbar$ and
$B=\sqrt{16E_{\rm{JE}}/E_{\rm{C}}}$ (solid lines) fit well the numerical
data of the tunnel junctions at large masses.}
\label{ab-coefficients}
\end{figure}

At $T \rightarrow 0$ escape from the potential wells is mainly due to 
quantum tunneling. The tunneling rate has the form, see for example
Refs.~\cite{Caldeira83,Larkin84},
\begin{equation}
\Gamma_{\rm{QT}} = A \exp{\left(-B\right)} \;\; .
\label{ab-rate}  
\end{equation}
In WKB approximation and neglecting damping, the coefficients
\begin{equation}
  A = A_0 \frac{2\epsilon_0}{\hbar} \sqrt{\frac{B}{2\pi}}
\end{equation}
and
\begin{equation}
  B = \frac{1}{e} \int{ \sqrt{C\left[E(\varphi)-\epsilon_0\right]}} d
\varphi \;\; .
\end{equation}
The integral runs over that phase space in which $E(\varphi)>\epsilon_0$.
The parameter $A_0 = \sqrt{60}$ for the cubic potential of Refs.~
\cite{Caldeira83,Larkin84}. (We did no succeed in deriving $A_0$, but
we hope that it does not depend strongly on the shape of the potential).
We have estimated the coefficients $A$ and $B$ both for tunnel and metallic
junctions (Fig. \ref{ab-coefficients}). Neglecting the exact shape of the
washboard potential, both $A$ and $B$ may be overestimated by about 20 per
cent. In the tunneling rate these corrections partly compensate.
Our numerical results strongly deviate from the analytical form when the 
mass becomes small. The zero-point energy is then quite high, the 
effective potential the particle has to tunnel through becomes small,
and the WKB approximation breaks down.

Caldeira and Leggett \cite{Caldeira83} and Larkin {\em et al.}
\cite{Larkin84} derived the rate
\begin{equation}
\Gamma_{\rm{QT}} = \gamma \frac{\omega_{\rm{p}}}{2\pi}
  \exp{\left(-\frac{14.4 E_{\rm{JE}}}{\hbar\omega_{\rm{p}}}
  \left[1+\frac{0.87}{Q}+...\right]\right)}
\label{qtrate}  
\end{equation}
with $\gamma \approx \sqrt{ 120\pi (14.4 E_{\rm{JE}}/\hbar \omega_{\rm{p}})}$
for tunneling out of a cubic potential.
Eq.~\ref{qtrate} holds exactly at weak damping, when the quality factor
$Q \gg 1$. The tunneling rate is considerably reduced at strong damping 
$Q \ll 1$. In what follows, and for ease of calculation, we assume the 
junctions are only weakly damped by including the term linear in $1/Q$. 

Note that for the sinusoidal current-phase relation and the washboard
potential (that is for $I\ll I_{\rm{c}}$) the prefactor in the exponent
is not 14.4 as for the cubic potential (that is for $I\approx I_{\rm{c}}$)
but $8\sqrt{2} \approx 11.3$. Since this is also a small correction, like
for the $A$ and the $B$ coefficients mentioned above, we will discuss our
experimental results using the above tunneling rate Eq.~\ref{qtrate}.
This also has the advantage of including possible corrections due to damping.

\begin{figure}
\hspace{15mm}\psfig{file=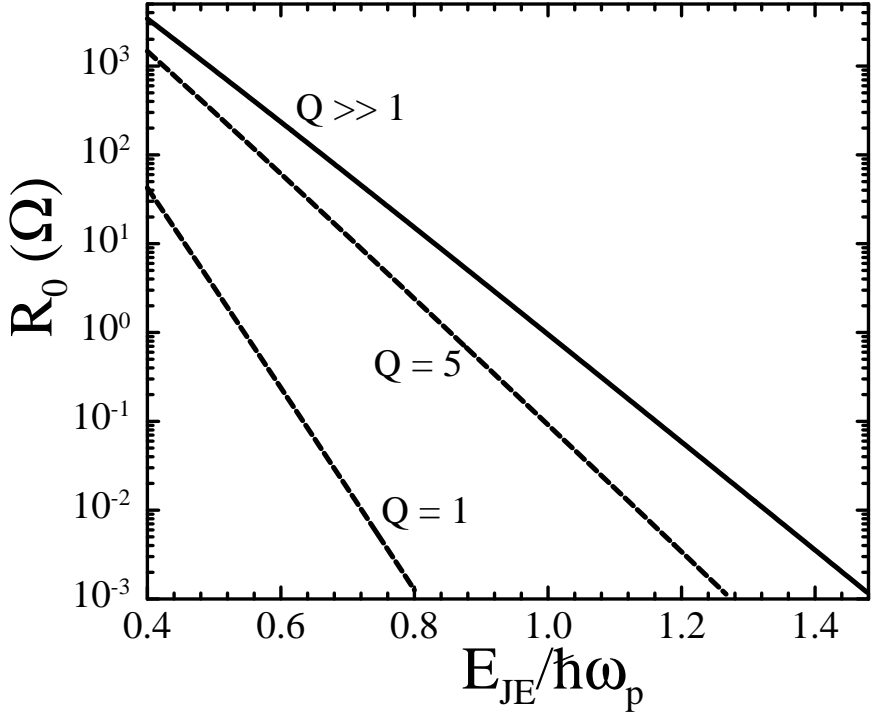,angle=270,width=150mm}\hfill
\caption[residual-plot-x]{\small \rm 
Residual resistance $R_0$ vs.~$E_{\rm{JE}}/\hbar\omega_{\rm{p}}$ at 
weak damping $Q \gg 1$, $Q = 5$, and $Q = 1$.}
\label{residual-plot}
\end{figure}

The asymmetry of the washboard potential due to a current flow yields a net 
phase slip rate $\dot{\varphi}$, that is the phase drifts. Tilting
the potential enhances the tunneling rate exponentially. Following Tinkham's
discussion \cite{Tinkham75} we get for the RCSJ model and Eq.~\ref{qtrate}
at weak damping
\begin{equation}
 \dot{\varphi}(x) \approx 
 4\pi \Gamma_{\rm{QT}} \sinh{\left(\frac{7.2\pi x E_{\rm{JE}}}
   {\hbar \omega_{\rm{p}}} \left[1+\frac{0.87}{Q}\right] \right)}
\;\; .
\label{phase-slip}
\end{equation}
Here we have used $E(x) \approx E_{\rm{JE}}(1-\pi x/2)$ and 
$\omega(x) \approx \omega_{\rm{p}}$, a good approximation as long as 
$x = I/I_{\rm{c}}^0 \le 0.5$. The phase slips result in a voltage drop 
$U = \hbar \dot{\varphi}/2e$. Consequently the differential resistance
\begin{equation}
 dU(I)/dI \approx R_0 \cosh{\left(\frac{3.6 \pi}{e\omega_{\rm{p}}}I
 \left[1+\frac{0.87}{Q}\right] \right)}
\label{dudi}
\end{equation}
with the residual resistance $R_0 = dU/dI(I=0)$ 
\begin{equation}
 R_0 \approx 
   1.8 \gamma R_{\rm{K}} \left[ 1+\frac{0.87}{Q}\right]
   \exp{\left(-\frac{14.4 E_{\rm{JE}}}{\hbar\omega_{\rm{p}}}
   \left[1+\frac{0.87}{Q}\right]\right)}
\;\; .
\label{residual}
\end{equation}
At $Q \gg 1$ one can directly read off the
$E_{\rm{JE}}/\hbar\omega_{\rm{p}}$-ratio from $R_0$ in
Fig.~\ref{residual-plot}. At strong damping, the phase
becomes more localized and $R_0$ being strongly reduced. The second-order
approximation of Eq.~\ref{dudi}
\begin{equation}
\frac{dU(I)}{R_0 dI} \approx 1 + 2
  \left[1+\frac{0.87}{Q}\right]^2
  \left(\frac{1.8\pi I}
  {e\omega_{\rm{p}}}\right)^2 
\label{derivative}
\end{equation}
allows to derive the plasma frequency $ \omega_{\rm{p}}$ and, through 
$R_0$, the coupling energy  $E_{\rm{JE}}$ from the $I(U)$ - characteristic 
at $I \ll I_{\rm{c}}^0$. In this approximation neglecting damping means to 
underestimate the plasma frequency by a factor of $\left[1+0.87/Q\right]$. 
The derivation of the Josephson coupling energy, on the other hand, is 
barely affected by damping because the correction factors almost cancel
each other.

To ensure tunneling from the lowest oscillator level at 
$\hbar\omega_{\rm{p}}/2$, the tilt of the washboard potential $h I/2e$ per 
period must be less than the energy difference of about 
$\hbar\omega_{\rm{p}}$ between the two lowest levels. At 
$I\ge I_{\rm{Z}} = e\omega_{\rm{p}}/\pi$ quantum tunneling populates
the second level, enhancing the total tunneling rate. This may be called
Zener tunneling of the phase. And because the phase diffuses faster than 
described by Eq.~\ref{phase-slip}, the differential resistance should
increase much stronger stronger than predicted by Eq.~\ref{dudi}.

The supercurrent is suppressed as soon as the phase diffuses or drifts
faster than $\omega_{\rm{p}}$ or at 
\begin{equation}
  1.8\left[ 1+\frac{0.87}{Q} \right] R_{\rm{K}} 
  \le R_0 \sinh \left(\frac{3.6\pi}{e \omega_{\rm{p}}}
  I \left[1+\frac{0.87}{Q}\right] \right)
\;\; .
\label{ic-of-r0}
\end{equation}
This relates the critical current 
\begin{equation}
I_{\rm{c}} =  \frac{I_{\rm{Z}}}{3.6\left[1+0.87/Q\right]} 
   \,{\rm{arcsinh}}\left(1.8\left[1+\frac{0.87}{Q}\right]
   \,\frac{R_{\rm{K}}}{R_0}\right)
\label{ic-phase}
\end{equation}
with the residual resistance and the plasma frequency. Here we have assumed
that the Q-factor does neither depend on $I$ nor on $U$. At $Q \gg 1$ the
critical current is larger than $I_{\rm{Z}}$ as long as 
$R_0 < 1.8 R_{\rm{K}}/\sinh (3.6) \approx 2.6\,{\rm{k}}\Omega$. 
Eq.~\ref{ic-phase} represents an upper limit for $I_{\rm{c}}$ 
because of neglecting Zener tunneling. Note that this equation is 
valid only when the actual critical current has been reduced already, that 
is at $x_{\rm{c}} \le 0.5$, while Eq.~\ref{reduced-ic}
represents a maximum for all $x$.

An estimate like in the previous section for the critical current yields for 
a $R_{\rm{N}} = 1\,{\rm{k}}\Omega$ junction a coupling energy of $E_{\rm{JE}}
\approx 0.2\,$meV and a plasma frequency of $\omega_{\rm{p}} \approx 0.57\,
({\rm{ps}}^{-1})$. Neglecting damping, the residual resistance due to phase 
diffusion Eq.~\ref{residual} amounts then to $R_0 \approx 
0.92\,{\rm{k}}\Omega$. Because of the exponential dependence in Eq. 
\ref{residual}, a $R_{\rm{N}} = 100\,\Omega$ junction has a tiny $R_0 
\approx 58\,\mu\Omega$ only. Thus the residual resistance rises steeply
within a very narrow resistance range.

\subsection{Bloch-wave oscillations}
To add a charge $q$ to the capacitance $C$ of the junction requires the 
Coulomb charging 
energy $E_{\rm{C}}=q^2/2C$. When $E_{\rm{C}} \ge E_{\rm{JE}}$, both phase 
and charge do no longer behave as classical variables, but like operators. 
Because of the periodic washboard potential (Eq.~\ref{jo-washboard}) the 
Josephson junction can then be described using Bloch waves 
\cite{Larkin84,Likharev85}. Their wave number is the quasi charge $k = q/2e$.
For a review see Ref.~\cite{Schoen90}.

Bloch-wave oscillations represent the periodic transfer of discrete Cooper 
pairs across the junction, which is recharged by the external current source
at a radial frequency of $\pi I/$e. 
These wave packages travelling in $\varphi$ space of the tilted washboard 
potential have a finite 'momentum' $<\partial \varphi/\partial t >$. This
implies a voltage drop, and the contact resistance becomes finite.
Theory predicts a region at finite bias 
current in which Bloch waves contribute negatively to the differential 
resistance \cite{Larkin84,Likharev85}, 
but the details of the spectra are difficult to derive unless 
the junctions are in the $E_{\rm{C}} \gg E_{\rm{JE}}$ limit. Since Bloch 
waves are suppressed by fluctuations, the quasi-particle resistance, that 
is the resistance $R_{\rm{qp}}$ of the RCSJ model, must be larger than about 
$R_{\rm{K}}/4 \approx 6.4\,{\rm{k}}\Omega$. 
It is therefore unlikely to observe them at metallic 
junctions if the quasi-particle resistance had the magnitude of $R_{\rm{N}}$. 
Intraband transitions can be induced when the external current is strong 
enough. This Zener tunneling of the quasi-charge requires 
$I\ge e\omega_{\rm{p}}/\pi$, like for Zener tunneling of the phase, see 
the previous section.

There are only few experimental reports on Bloch-wave oscillations at 
Josephson junctions, e.~g.~Ref.~\cite{Kuzmin91}. In those experiments  
Bloch oscillations have been identified by a systematic coincidence between 
structures of the $I(U)$ - characteristic at certain applied DC currents
and the frequency of an external microwave excitation.

\subsection{Andreev reflection}
Additional information on the properties of the junction comes from 
Andreev reflection at finite voltages. If the potential difference is 
$|U| > \Delta/e$, quasi-particles from one side of the junction can enter 
the other one by Andreev reflection, and return a quasi-hole. At high bias 
voltages this additional hole or excess current amounts to $I_{\rm{ex}} = 
8\Delta/3 e R_{\rm{N}}$ and $I_{\rm{ex}} = \left(\pi^2-4\right) 
\Delta/ 4 e R_{\rm{N}}$ in the clean and the dirty limit, respectively 
\cite{Blonder82,Artemenko79}. 

If $|U| < \Delta/e$ the quasi-hole returns at an energy still inside the 
gap, and it is reflected again as a quasi-particle. Since the charge 
carriers gain an energy $e|U|$ per crossing the junction, $M=2\Delta/e|U|$
($M=1, 2, 3, ...$) successive reflections are required to overcome the gap.
At $|U|=2\Delta/eM$ the conductance increases stepwise. This multiple Andreev
reflection becomes more pronounced at the presence of normal 
quasi-particle reflection at the interface, that is at $D<1$. Normal
quasi-particle reflection strongly reduces the excess current, because
both the quasi-particle and the quasi-hole have to cross the interface.

The Andreev-reflection excess current can also be reduced due to a finite
quasi-particle lifetime $\tau$. For normal-superconducting junctions,
lifetime effects are taken into account by the modified BTK theory
\cite{Plecenik94}, applying the Dynes' model. At superconducting junctions,
lifetime effects are expected to suppress structures due to multiple Andreev
reflection, see for example Refs.~\cite{Cuevas96,LevyYeyati96}.

As soon as phase diffusion becomes efficient and $R_0$ finite, it can be
difficult to distinguish clearly between the Josephson supercurrent and the
Andreev-reflection hole current. At large voltages $|U|\gg \Delta/e$ and
beyond the critical current, however, the Andreev-reflection process is not
affected by the capacity of the RCSJ model. Therefore the literature value
of $I_{\rm{ex}}$ has to be expected. This has been verified previously on
normal-superconducting junctions \cite{Gloos96-auin2}.

\begin{figure}
\hspace{40mm}\psfig{file=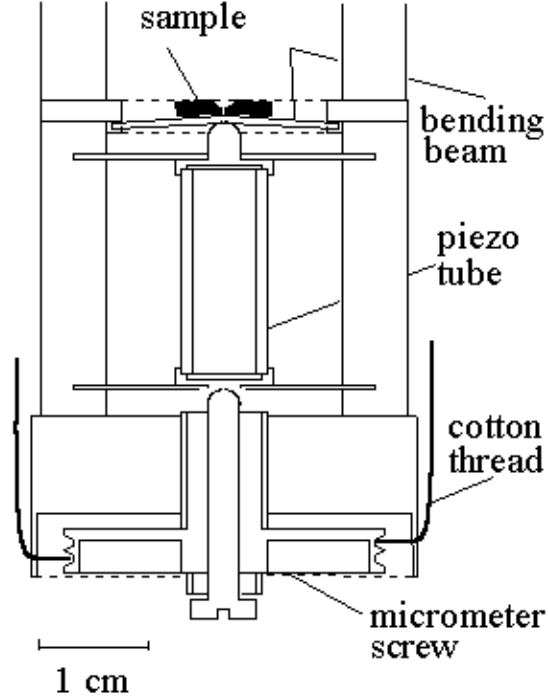,width=80mm}
\hfill
\\[20pt]
\caption[setup-x]{\small \rm 
Our break-junction device. The sample is broken by turning the micrometer
screw using two flexible cotton threads. This screw is also used for coarsely
adjusting the contact. Fine adjustment is achieved by applying a voltage
at the piezo tube.}
\label{setup}
\end{figure}

\begin{figure}
\hspace{40mm}\psfig{file=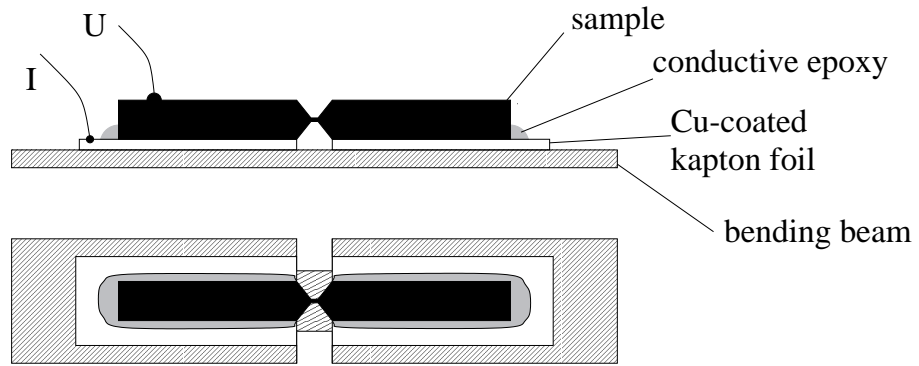,width=120mm}
\hfill
\\[20pt]
\caption[beam-x]{\small \rm 
The sample on the bending beam. This gold-plated beam is cut from copper 
alloy. Only one of the current and one of the voltage leads are shown.}
\label{beam}
\end{figure}

\section{Experiment}

We prepared metallic point contacts between two bulk pieces of the BCS-type 
superconductors lead ($T_{\rm{c}}=7.2\,$K), indium (3.4 K), aluminum (1.15 K), 
and cadmium (0.54 K) using mechanical-controllable break junctions.
Fig.~\ref{setup} shows our setup and Fig.~\ref{beam} the sample on the
bending beam. The samples, $\sim\,$1 mm diameter wires, were 
broken at a predefined notch in the ultra-high vacuum region of the cold 
refrigerator. This avoids oxidation of the interfaces and ensures the 
junctions to consist of pure metal. 
The lateral contact size and, thus, the normal-state resistance could be 
adjusted in situ by a micrometer screw, driven mechanically by a  
pulley-and-rope and a piezo tube. The current-voltage 
characteristic and the differential resistance were recorded in the standard 
four-terminal mode with current biasing.

Twisted pairs of wires are used for the current and voltage leads. They
are properly shielded. Simple LRC filters at the mixing chamber protect the
sample from low-frequency ($\sim$MHz) noise. We have no copper-powder filter
as described in Refs.~\cite{Muller92,Muller94,Peters95,vanderPost97}.

We start the experiments with the contact resistance set to about 
$1 - 3\,{\rm{m}}\Omega$ at room temperature. This corresponds to a
contact radius of about $10\,\mu$m. The residual resistance ratio of the 
material in the contact region was then observed if possible while cooling
down. We obtained as lower bound of the electronic mean free path 
$l \ge 70\,$ nm (Pb), $310\,$nm (In), $545\,$nm (Al), and $75\,$nm (Cd). 
The junctions investigated here have $R_{\rm{N}} \ge 1\,\Omega$. They are 
therefore ballistic and have, according to Sharvin's resistance formula
\cite {Sharvin65}, that is Eq.~\ref{ballistic} at $D=1$, a contact area of
about $665\,{\rm{nm}}^2/R_{\rm{N}}(\Omega)$. 
The bulk samples had residual resistance ratios of about 
7570 (Pb), 4760 (In), 1270 (Al), and 1430 (Cd). This corresponds to 
electronic mean free paths  of about 18 $\mu$m (Pb), 29 $\mu$m (In), 
19 $\mu$m (Al), and  12 $\mu$m (Cd).
Unless otherwise specified, the following experimental data have been
obtained at low temperatures of $T\approx 50\,$mK.

\begin{figure}
\hspace{15mm}\psfig{file=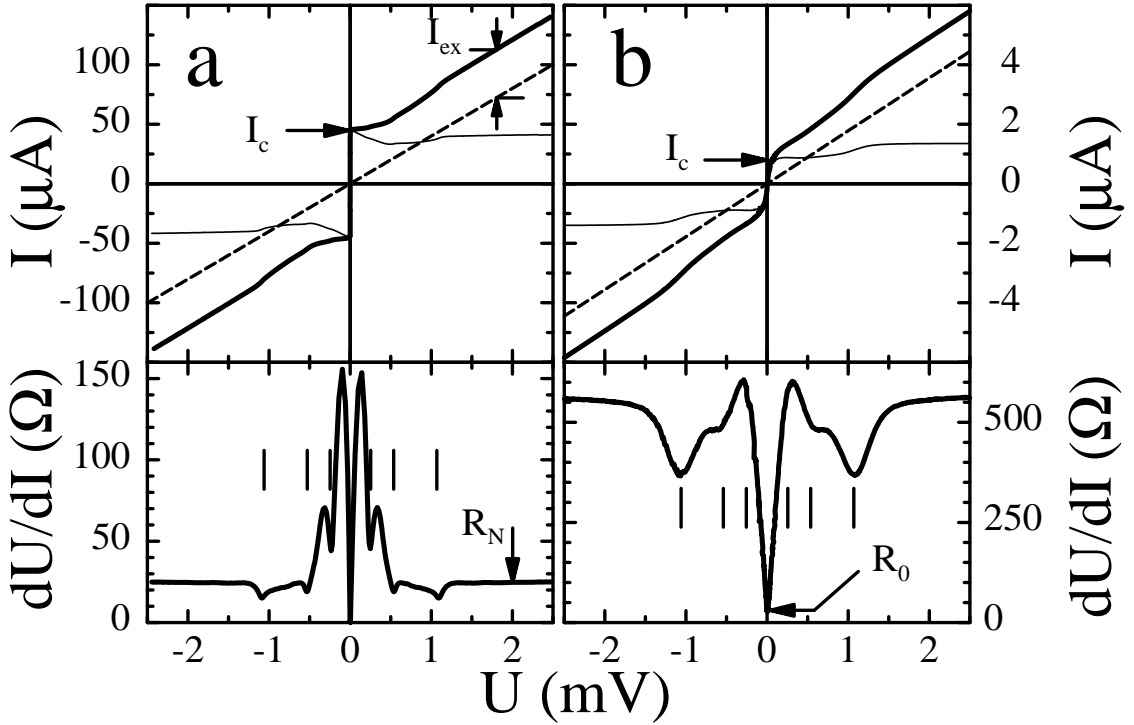,angle=270,width=150mm}
\caption[spec-x]{\small \rm 
$I$ vs.~$U$ and $dU/dI$ vs.~$U$ of a) a low-resistance and b) a 
high-resistance In junction at $T = 50\,$mK(thick solid lines). 
Arrows define the various parameters used. The vertical bars mark the 
superconducting energy gap $2\Delta/e$ and its subharmonics. The dashed
lines are the normal-state $I(U)$ characteristics. The thin solid lines 
represent the additional current due to the Josephson effect and the 
Andreev-reflection excess current.}
\label{defspec}
\end{figure}

\section{Results and discussion}

The $I(U)$ - characteristics as well as the differential resistance $dU/dI$ 
could be recorded over the full superconducting anomaly without excessive 
heating for junctions with $R_{\rm{N}} \ge 1\,\Omega$. Few junctions with
$R_{\rm{N}} \approx 1 - 10\,\Omega$ were hysteretic, either due to heating
or, according to the RCSJ model, due to being weakly damped.
The self-magnetic field at the junctions could also contribute to hysteresis,
but this has to be expected at much smaller contact resistances
$R_{\rm{N}} \ll 1\,\Omega$. For those hysteretic junctions the critical
current was extracted from that branch of the $I(U)$ - characteristic with
increasing current.

\begin{figure}
\hspace{20mm}\psfig{file=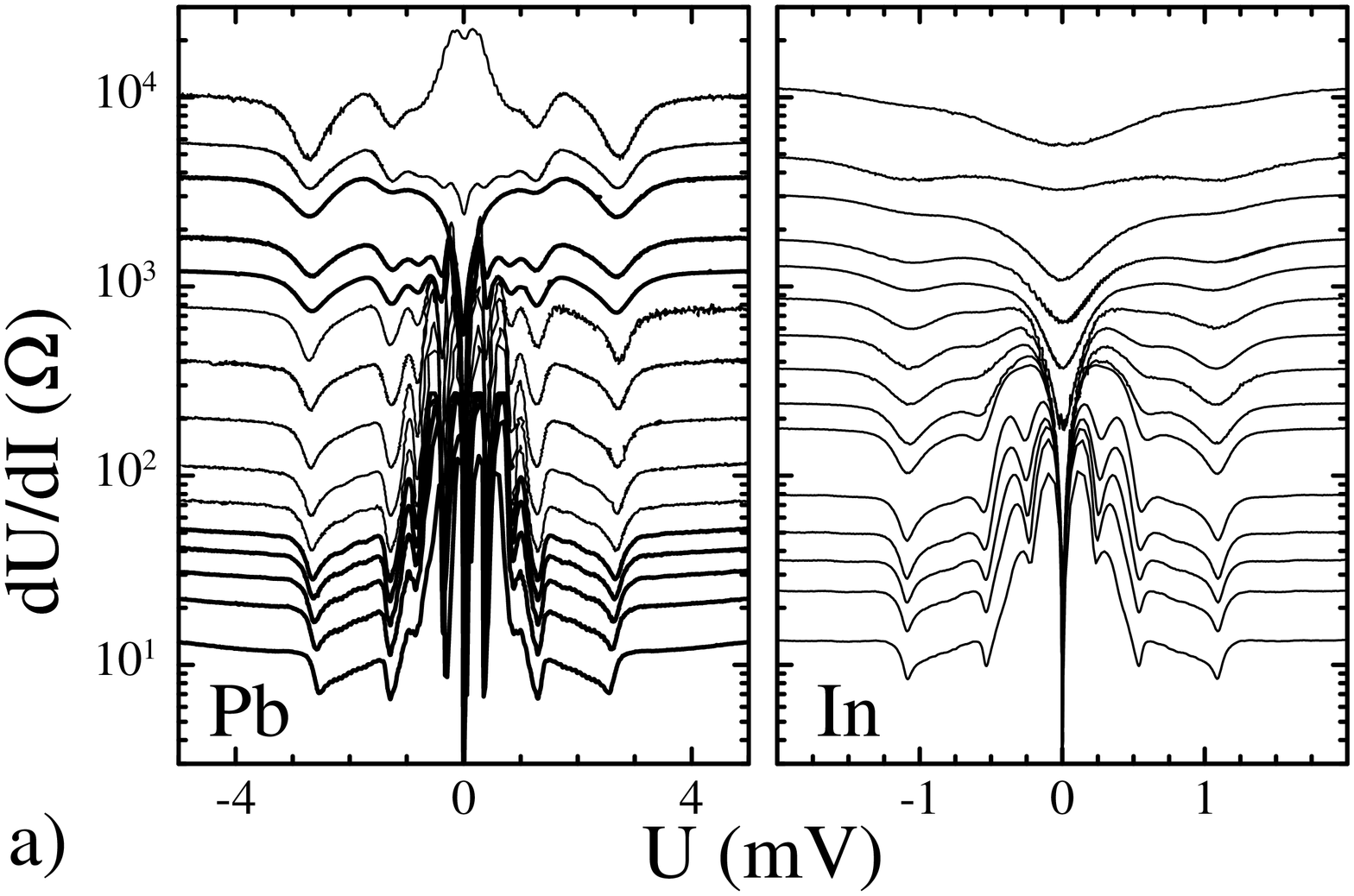,angle=270,width=120mm}\hfill

\hspace{20mm} \psfig{file=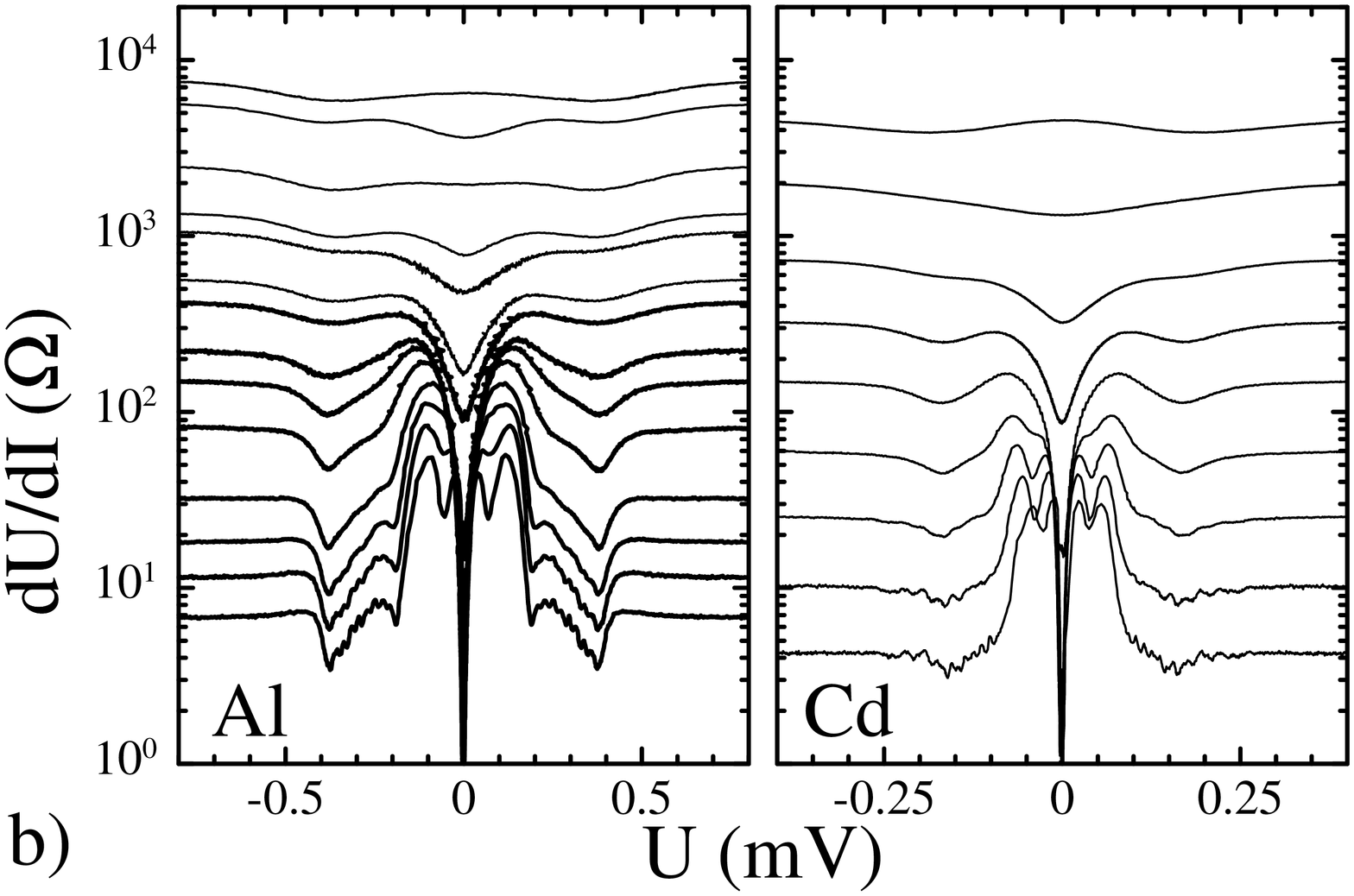,angle=270,width=120mm}\hfill
\caption[spec-x]{\small \rm 
Typical spectra $dU/dI$ vs.~$U$ of a) Pb and In and b) Al and Cd junctions at 
$T = 50\,$mK.}
\label{spectra}
\end{figure}

Fig.~\ref{defspec} shows two typical $I(U)$ - characteristics and $dU/dI$ 
vs.~$U$ spectra as example and for defining the various parameters. A number 
of selected spectra are shown in Fig.~\ref{spectra}. They look quite similar
for all four superconductors. There are distinct structures due to 
superconductivity at small $R_{\rm{N}}$, including an unresolvably small
$R_0$. These structures diminish at larger
resistances, but their positions on the voltage axis do barely change. 
Finally, in the resistance  range around $10\,{\rm{k}}\Omega$ we observe, as
expected, the transition to the tunneling regime, with zero-bias maxima
instead of minima. The contact resistance $R_{\rm{N}}$ up to which well
pronounced superconducting anomalies could be seen was the higher the larger
the superconducting gap was.

\begin{figure}
\hspace{20mm}\psfig{file=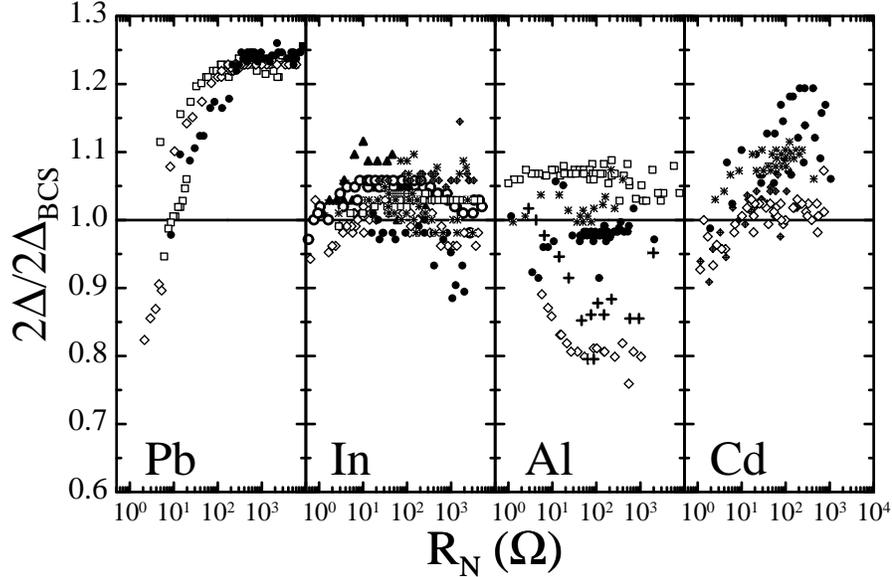,angle=270,width=120mm}
\caption[gap-x]{\small \rm 
Experimental superconducting gap $2\Delta$ at $T=50\,$mK vs.~$R_{\rm{N}}$, 
normalized to the BCS value $2\Delta_{\rm{BCS}}=3.52\,k_{\rm{B}}T_{\rm{c}}$.
Identical symbols are used in Figs.~\ref{gap} - \ref{analysis-c} for the 
same pieces of sample.} 
\label{gap}
\end{figure}

The gap structure at $U=2\Delta/e$ and its subharmonics
due to multiple Andreev reflection agreed well
with the BCS value $2\Delta_{\rm{BCS}} = 3.52\,k_{\rm{B}}T_{\rm{c}} 
=$ 1.05 meV, 352 $\mu$eV, and 165 $\mu$eV of In, Al, and Cd, 
respectively (Fig.~\ref{gap}). 
Strong-coupling Pb had the width enhanced by about 25 per cent with respect 
to $2\Delta_{\rm{BCS}}=$ 2.20 meV (Fig.~\ref{gap}).
Why the Pb junctions did not have the right energy gap at small
resistances we do not know yet. Heating effects or the suppression of 
superconductivity by the self-magnetic field can be excluded because 
the other superconductors behave regularly (at least for junctions with
$R_{\rm{N}}\ge 10\,\Omega$).
Some of the Al junctions had lower $T_{\rm{c}}$ down to 0.85 K, while some 
of the Cd junctions had higher $T_{\rm{c}}$ up to 0.65 K. This is probably 
due to disorder and stress in the 
contact region and resulted in smaller or larger gaps, respectively.

\begin{figure}
\hspace{20mm}\psfig{file=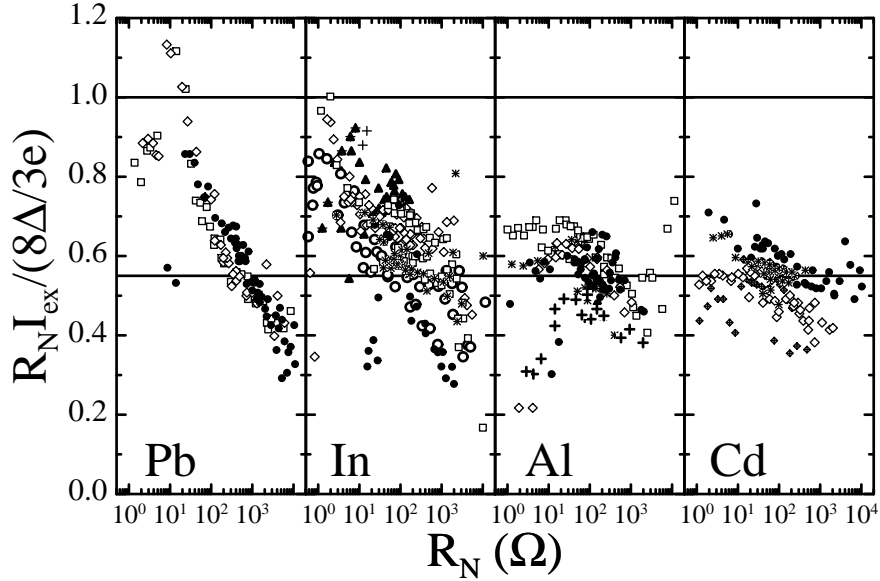,angle=270,width=120mm}
\caption[rniex-x]{\small \rm 
Experimental $R_{\rm{N}}I_{\rm{ex}}$ vs.~$R_{\rm{N}}$ at $T=50\,$mK, 
normalized to the clean-limit (KO2) value $8\Delta/3e=$ 3.67, 1.40,
0.47, and 0.22 mV for Pb, In, Al, and Cd, respectively. 
Solid lines indicate the KO2 ($=1.0$) and 
KO1 ($=0.55$) value.}
\label{rniex}
\end{figure}

\subsection{Andreev-reflection excess current}
At small $R_{\rm{N}}\approx 1\,\Omega$, the excess current $I_{\rm{ex}}$ was
near the clean (KO2) limit for Pb (using the experimental $2\Delta =
2.75\,$meV) and In, and near the dirty (KO1) limit for Al and Cd 
(Fig.~\ref{rniex}). This trend could be consistent with the superconducting
coherence length $\xi \propto 1/\Delta$ being shorter for Pb and In than
for Al and Cd. However, according to the above bulk resistivity data
all four superconductors should have been in the clean limit.

At larger $R_{\rm{N}}$ and towards the metallic-tunneling transition at
$R_{\rm{K}}/2$, the excess current of the Pb and In junctions decreased but
remained finite, settling around the dirty limit.
This reduction may have several reasons:
First, small contacts could be stronger distorted than large ones.
Second, the junctions could have a finite quasi-particle density of states
due to lifetime effects. Using the modified BTK theory \cite{Plecenik94},
a reduction of the excess current by a factor two requires
a lifetime parameter not larger than $\Gamma = \hbar/\tau \approx \Delta/5$.
It agrees roughly with $\Gamma \approx 50\,\mu$eV, estimated using  the bulk
residual resistivity of the samples. This estimate would be supported by the
clearly resolved multiple Andreev reflection anomalies of Pb and In, while
Al and Cd show more washed out and smeared structures.
Third, normal quasi-particle reflection. Since multiple Andreev
reflection is observed, the transmission coefficient $D$ has to be at
least somewhat less than 1. An excess current of (on the average)
not less than half the maximum possible one, makes the lower bound of the
transmission coefficient $D \ge 0.5$. On the other hand, all
four superconductors show the transition from vacuum tunneling to metallic
conduction at around $R_{\rm{K}}/2$, indicating $D\approx 1$.

\begin{figure}
\hspace{20mm}\psfig{file=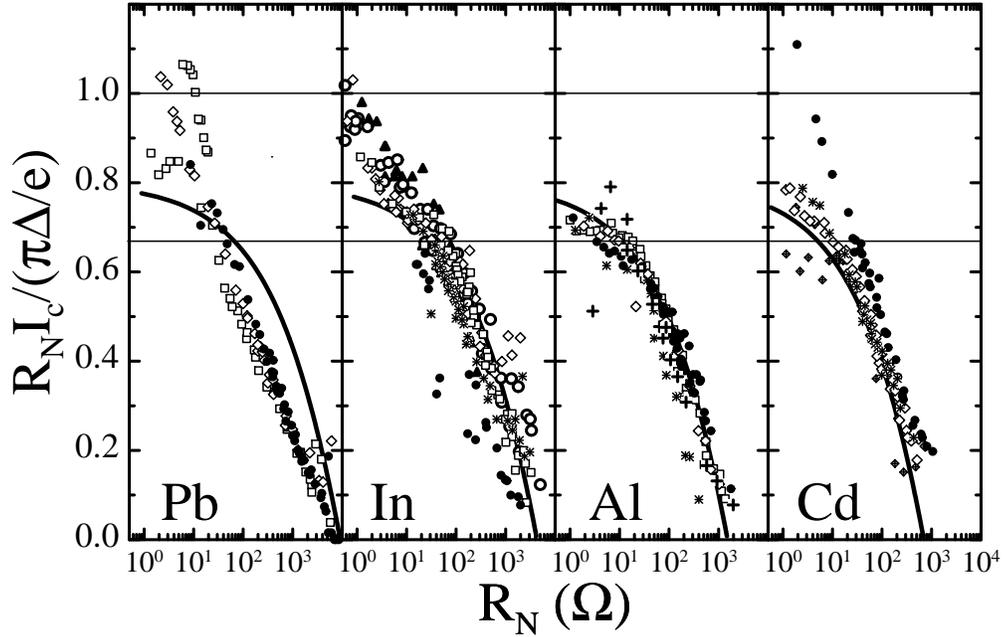,angle=270,width=135mm}\hfill
\caption[rnic-x]{\small \rm 
Experimental $R_{\rm{N}}I_{\rm{c}}$ vs.~$R_{\rm{N}}$ at $T=50\,$mK, 
normalized to the clean-limit (KO2) value $\pi\Delta/e=$ 4.32, 1.65,
0.55, and 0.26 mV for Pb, In, Al, and Cd, respectively.
Solid lines indicate the KO2 ($=1.0$) and the KO1 ($=0.66$) result as well as 
that due to the zero-point fluctuations (Eq.~\ref{icrn}) at $C=0.05\,$fF and
the abscissa fixed at 0.8, the best fit for aluminum.}
\label{rnic}
\end{figure}

\subsection{Critical current}
At small $R_{\rm{N}}\approx 1\,\Omega$, the critical current $I_{\rm{c}}$
(Fig.~\ref{rnic}) was near the clean (KO2) limit for Pb and In, and near the
dirty (KO1) limit for Al and Cd, like the excess current. But
$R_{\rm{N}} I_{\rm{c}}$  decreased continously at larger contact resistances 
and vanished well below $R_{\rm{K}}/2$. This agrees with previous
observations by others \cite{Muller92,Muller94,Peters95,vanderPost97}.

Like for the excess current, we would have expected the critical current or
the product $R_{\rm{N}} I_{\rm{c}}$ to be reduced, but to the dirty limit
only. Assuming a minimum $D \approx 0.5$ for normal reflection from the
Andreev-reflection excess current and applying the current-phase
relationship of Eq.~\ref{cpr}, the critical current can be reduced by
almost a factor two  near the Ambegaokar-Baratoff value
$\pi\Delta/2eR_{\rm{N}}$.
Thus the reduction of $R_{\rm{N}} I_{\rm{c}}$ to zero requires an explanation
that does neither affect the superconducting gap nor the excess current.
The RCSJ model could solve this problem: the high-resistance part of
$R_{\rm{N}}I_{\rm{c}}$ can be described by Eq.~\ref{icrn} and a small 
capacitance of $C\approx 0.05\,$fF. The fit is nearly perfect for Al, but
deviates for the other superconductors. Especially the nearly straight line
of Pb in the semi-log plot astonishes. These deviations from the theoretical
curve at constant $C$ require additional processes, for example a variation
of the intrinsic $R_{\rm{N}}I_{\rm{c}}^0$ at the junction, a varying
amount dissipation, or a varying capacitance.

\begin{figure}
\hspace{20mm}\psfig{file=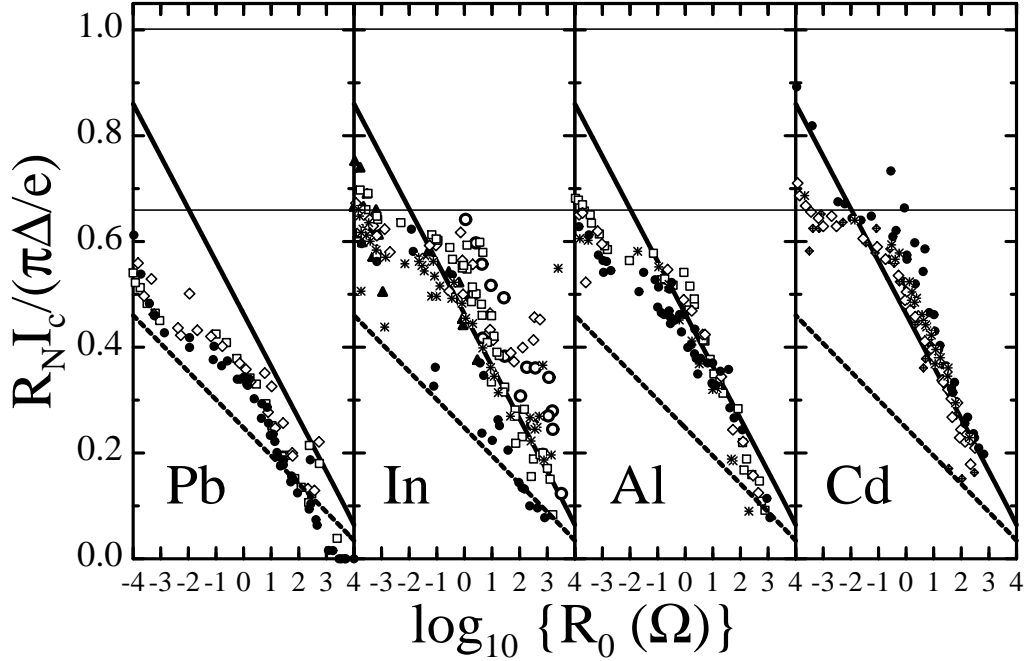,angle=270,width=140mm}\hfill
\caption[ic-r0-x]{\small \rm 
Experimental $R_{\rm{N}}I_{\rm{c}}$ vs.~$R_{\rm{0}}$ at $T=50\,$mK, 
normalized to the clean-limit (KO2) value $\pi\Delta/e$. 
Indicated are the KO2 ($=1.0$) and the KO1 ($=0.66$) result as well as that 
due to phase diffusion Eqs.~\ref{ic-phase} and \ref{ic-of-r0-kappa} at 
a lead capacitance of $\kappa = 3.3\,$pF/m and a quality factor $Q \gg 1$
(solid lines) and $Q=1$ (dashed lines).}
\label{ic-r0}.
\end{figure}

\begin{figure}
\hspace{20mm}\psfig{file=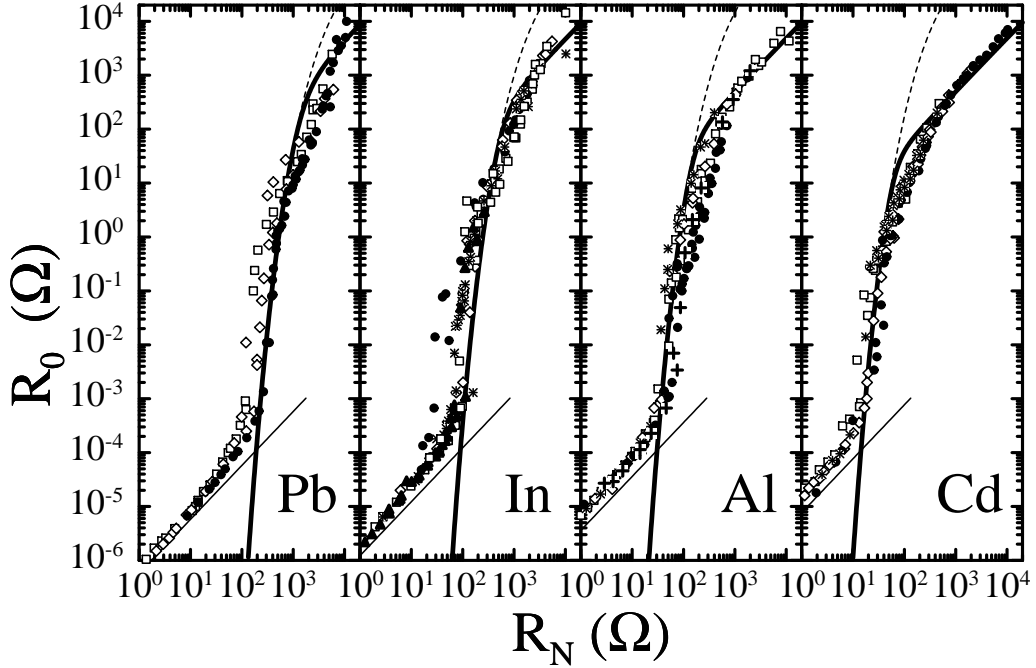,angle=270,width=140mm}\hfill
\caption[R0-RN-X]{\small \rm 
Residual resistance $R_0$ vs.~$R_{\rm{N}}$ at $T=50\,$mK. Straight lines 
indicate the detection limit, dashed lines are $R_0$ of Eq.~\ref{residual} 
at weak damping $Q\gg 1$, using the clean-limit $I_{\rm{c}}^0$ and 
$C=0.05\,$fF.  Solid lines take into account $R_{\rm{N}}/2$ in parallel.}
\label{R0-RN}
\end{figure}

Fig.~\ref{ic-r0} shows an alternative plot of $R_{\rm{N}} I_{\rm{c}}$ vs.~
residual resistance $R_0$. Again a systematic decrease is observed when 
$R_0$ increases. But this time, all four superconductors behave 
similarly. The fit through the data points takes into account the dynamic
capacitance of the junctions as discussed later.

\subsection{Residual contact resistance}
At low $R_{\rm{N}}$ the residual contact resistance
$R_0=dU/dI(I=0)$ was unresolvably small. Here the experimental resolution 
results from the $\sim\,2$ nV detection limit at a maximum current of $I_c$.
Towards larger $R_{\rm{N}}$ the residual resistance emerged out of the
detection limit of $\sim 2\,{\rm{nV}}/I_{\rm{c}}$, first rising steeply,
and then approaching $\sim R_{\rm{N}}/2$ (Fig.~\ref{R0-RN}). Again, this
behaviour agrees qualitatively with previous observations by others
\cite{Muller92,Muller94,Peters95,vanderPost97}.

\begin{figure}
\hspace{20mm}\psfig{file=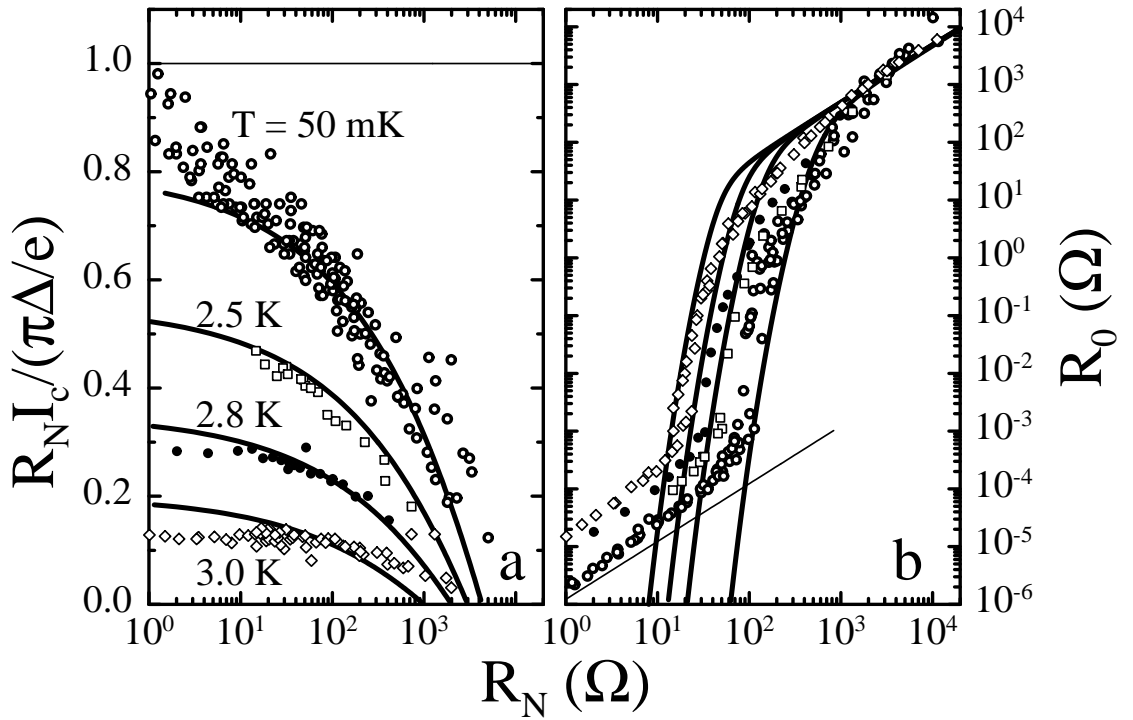,angle=270,width=150mm}\hfill
\caption[R0-temp-x]{\small \rm 
a) Experimental $R_{\rm{N}}I_{\rm{c}}$ vs.~$R_{\rm{N}}$ of In at the
indicated temperatures. The data are normalized to the clean-limit (KO2) 
value $\pi\Delta/e$ at $T \rightarrow 0$. 
Solid lines result from the zero-point fluctuations Eq.~\ref{icrn} at 
$C=0.05\,$fF.
b) Residual resistance $R_0$ vs.~$R_{\rm{N}}$. Temperatures and symbols 
as in a). The straight line represents the detection limit for the 
$T=50\,$mK data, solid lines are $R_0$ of Eq.~\ref{residual} using the 
clean-limit $I_{\rm{c}}^0(T)$ and $C=0.05\,$fF, and taking into account 
the asymptotic $R_{\rm{N}}/2$.}
\label{R0-temp}
\end{figure}

Raising the temperature, for example to $T=2\,$K for indium, did barely 
alter the results. This indicates that the junctions are not thermally 
activated but are in the regime of quantum tunneling. Only near $T_{\rm{c}}$, 
when the intrinsic $(R_{\rm{N}} I_{\rm{c}}^0)$ decreases strongly, we also 
found the transition to a finite $R_0$ at smaller $R_{\rm{N}}$ 
(Fig.~\ref{R0-temp}).

The data in Fig.~\ref{R0-RN}, especially the steep rise of $R_0(R_{\rm{N}})$, 
can well be described by  Eq.~\ref{residual} using a constant 
$C \approx 0.05\,$fF as for the critical current in the previous section, 
the clean-limit $I_{\rm{c}}^0$, and weak damping. 
Deviations at large $R_{\rm{N}}$ have to be expected:  first, the above
analysis does not apply when the experimental $R_{\rm{N}}I_{\rm{c}} = 0$.
Second, in the dissipative state the total resistance of the junction can 
not exceed $R_{\rm{N}}$. It must be less than $R_{\rm{N}}/2$ if one 
includes Andreev reflection, that has not been considered by Eq.~\ref{qtrate}. 
Combining phenomenologically these two different regimes, that is $R_0$ 
of Eq.~\ref{residual} and $R_{\rm{N}}/2$, by substituting 
$1/R_0 \rightarrow 1/R_0 + 2/R_{\rm{N}}$, 
fits the data of all four superconductors.

\subsection{Plasma frequency and Josephson coupling energy}
\begin{figure}
\hspace{20mm}\psfig{file=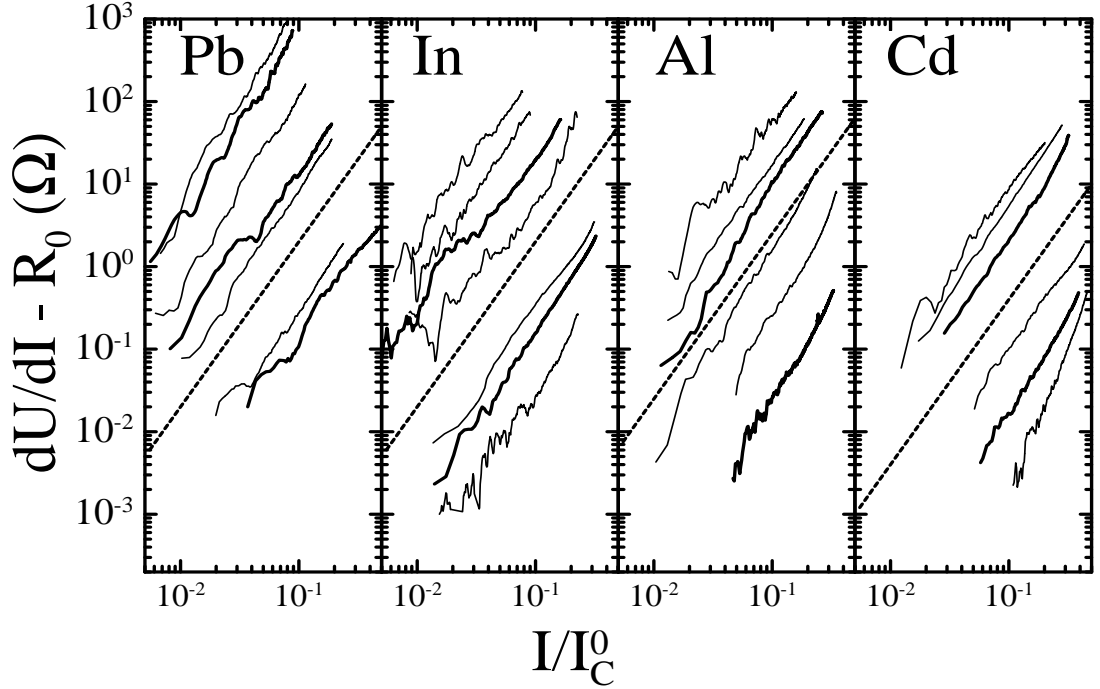,angle=270,width=150mm}\hfill
\caption[dudi-i2-x]{\small \rm 
Typical $dU/dI-R_0$ vs.~$I/I_{\rm{c}}^0$ at $T = 50\,$mK. Here
$I_{\rm{c}}^0$ is the clean-limit value derived from $R_{\rm{N}}$.
The dashed lines have a slope of 2 in the double-log plot.}
\label{dudi-i2}
\end{figure}

According to Eqs.~\ref{dudi} - \ref{derivative} the plasma frequency as well 
as the coupling energy of each contact with a sufficiently large residual 
resistance can be derived from the differential resistance at small currents. 
We find indeed a linear dependence between $dU/dI$ and $I^2$ 
(Fig.~\ref{dudi-i2}). Its slope divided by $R_0$ yields directly the
plasma frequency $\omega_{\rm{p}}$, assuming weak damping $Q \gg 1$. This 
in turn is used to extract $E_{\rm{JE}}$ from $R_0$. The contacts that can
be analyzed this way have residual resistances of $\sim 0.1\,\Omega$ to 
$\sim 1\,{\rm{k}}\Omega$, and according to Fig.~\ref{residual-plot} their
$E_{\rm{JE}}/\hbar\omega_{\rm{p}}$ varies from about 1.2 to 0.5. 

\vspace*{-15mm}
\begin{figure}
\hspace{20mm}\psfig{file=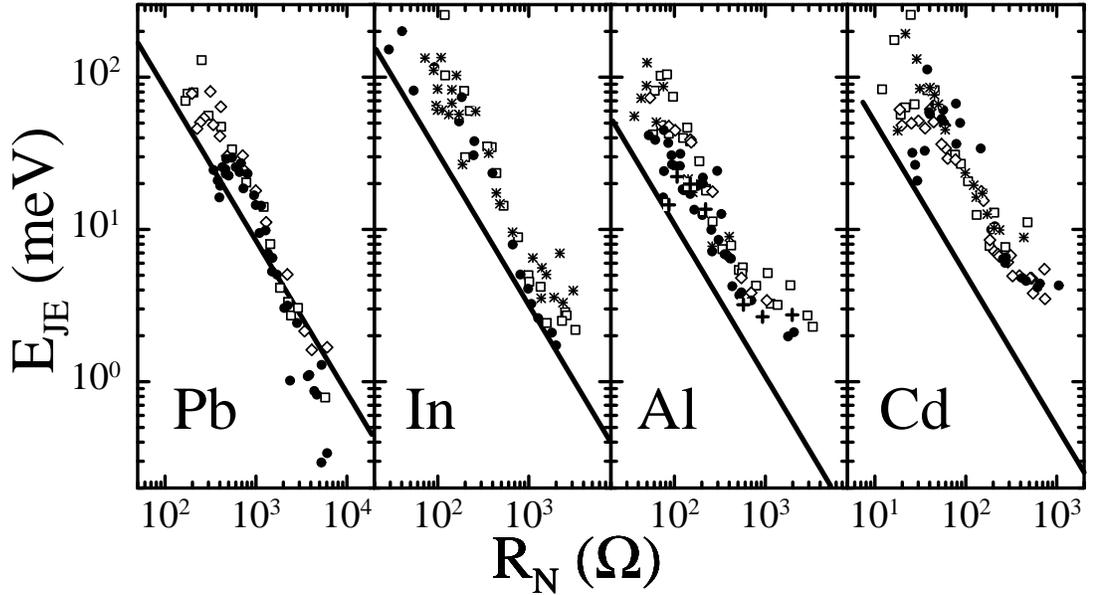,angle=270,width=150mm}\hfill
\caption[analysis-eje-x]{\small\rm
 Josephson coupling energy $E_{\rm{JE}}$ vs.~$R_{\rm{N}}$ derived from  
 $dU/dI$ at $I \rightarrow 0$ and at $T=50\,$mK, assuming weak damping 
 $Q\gg 1$. Solid lines are $E_{\rm{JE}}=\Delta R_{\rm{K}}/4R_{\rm{N}}$ 
 in the clean limit.}
\label{analysis-eje}
\end{figure}

\begin{figure}
\hspace{20mm}\psfig{file=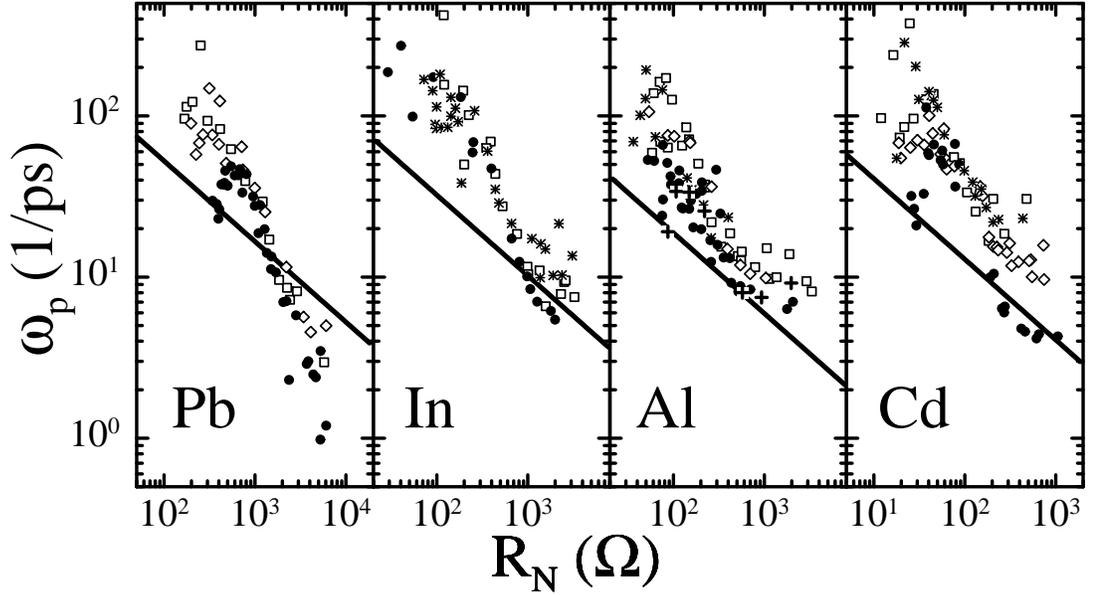,angle=270,width=150mm}\hfill
\caption[analysis-wp-x]{\small\rm
 Plasma frequency $\omega_{\rm{p}}$ vs.~$R_{\rm{N}}$ derived from 
 $I(U)$ at $I \rightarrow 0$ and at $T=50\,$mK. Solid lines 
 are $\omega_{\rm{p}}=\sqrt{2\pi\Delta/\hbar R_{\rm{N}} C}$ at 
 $C=0.05\,$fF and in the clean limit.}
\label{analysis-wp}
\end{figure}

\begin{figure}
\hspace{20mm}\psfig{file=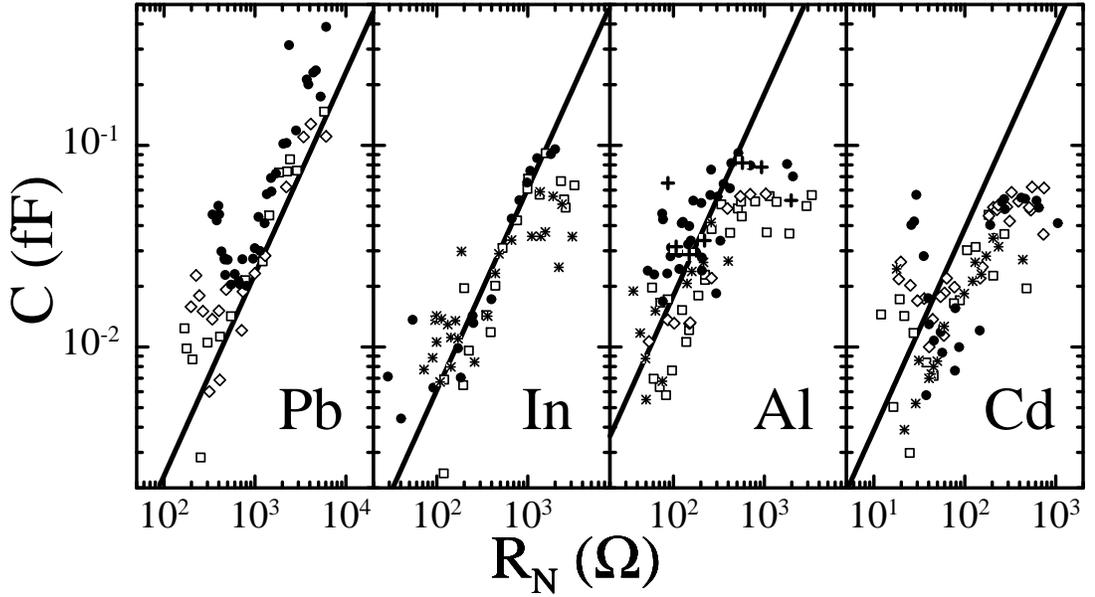,angle=270,width=150mm}\hfill
\caption[analysis-c-x]{\small\rm
 Capacitance of the junctions $C$ derived from Eq.~\ref{exp-cap} and the 
 data of Figs.~\ref{analysis-eje} and \ref{analysis-wp} vs.~$R_{\rm{N}}$.
 Solid lines represent $C = 0.1\,{\rm{fF}}/I_{\rm{c}}^0 (\mu{\rm{A}})$ and 
 the clean-limit $I_{\rm{c}}^0$.}
\label{analysis-c}
\end{figure}

In contrast to the strong reduction of the experimental $I_{\rm{c}}$ in 
Fig.~\ref{rnic}, the coupling energy $E_{\rm{JE}}$ at $I \rightarrow 0$ in 
Fig.~\ref{analysis-eje} is  slightly larger than expected from the 
clean-limit $I_{\rm{c}}^0$. 
The plasma frequency $\omega_{\rm{p}}$ corresponds to a capacitance of order 
$C = 0.05\,$fF only at large contact resistances (Fig.~\ref{analysis-wp}). 
Deviations at small $R_{\rm{N}}$ can not be explaind by a reduced escape 
rate due to stronger damping, because including damping would lead to an even 
larger $\omega_{\rm{p}}$ and not to a smaller one. Thus we are forced to 
conclude that the capacitance $C$ is not a constant. The effective 
capacitance of the junctions 
\begin{equation}
  C \approx \frac{4e^2 E_{\rm{JE}}}{\hbar^2 \omega_{\rm{p}}^2}
\label{exp-cap}
\end{equation}
becomes $C \approx 0.1\,{\rm{fF}}/I_{\rm{c}}^0(\mu{\rm{A}})$, depending
on the intrinsic critical current (Fig.~\ref{analysis-c}). $C$ does 
not depend on the geometrical cross sectional area of the junctions which 
is inversely proportional to $R_{\rm{N}}$. We will show below that it is the 
above relationship  with $I_{\rm{c}}^0$ that has indeed to be expected. 

\vspace*{-15mm}
\vspace*{-15mm}

For the junctions with finite $R_0$, the
$E_{\rm{JE}}/\hbar\omega_{\rm{p}}$-ratio is always close to one, that is
these junctions are very susceptible to quantum fluctuations and phase
diffusion. And it requires only slight variations of this ratio to produce
the steep rise of $R_0(R_{\rm{N}})$ because of the large prefactor 14.4 in
the exponential function of Eq.~\ref{residual}.

If one assumes $R_{\rm{qp}}=R_{\rm{N}}$, the damping factors
$Q = 8\pi R_{\rm{qp}}E_{\rm{JE}}/R_{\rm{K}} \hbar\omega_{\rm{p}}
\approx R_{\rm{qp}}({\rm{k}}\Omega)$ may justify our data analysis using
the $Q \gg 1$ limit for the tunneling rate $\Gamma_{\rm{QT}}$ and neglecting
damping for Pb junctions at $R_{\rm{N}} > 1\,{\rm{k}}\Omega$ but not for Cd,
despite both superconductors show similar results. The question arises
whether it is correct to describe the damping term of these ballistic
junctions by their normal-state resistance $R_{\rm{N}}$. Our experiments
seem to indicate that the true quasi-particle resistance of these ballistic
Josephson junctions at $I \ll I_{\rm{c}}^0$ is much larger than $R_{\rm{N}}$.
Such an enhanced $R_{\rm{qp}}$ has to be expected according to
Eq.~\ref{qp-damping}, it finds its natural explanation in the gap of
the quasi-particle density  of states, that is there are no quasi-particles
at all (the clearly resolved multiple Andreev-reflection signal shows that
lifetime effects are negligible). Damping can then only be due to losses
of the high-frequency electromagnetic field around the junction
in the sample and in the metallic part of the setup.

Simultaneously, the absence of hysteretic $I(U)$-characteristics that would
be typical for weakly damped junctions can be explained by strong phase
diffusion or drift at $I_{\rm{c}}$ that hinders trapping of the phase. One
may also speculate whether the Q factor near $I_{\rm{c}}$ is smaller than
that at $I\rightarrow 0$, due to the frequency dependence of the
electromagnetic field generated by the alternating Josephson supercurrent.

\begin{figure}
\hspace{20mm}\psfig{file=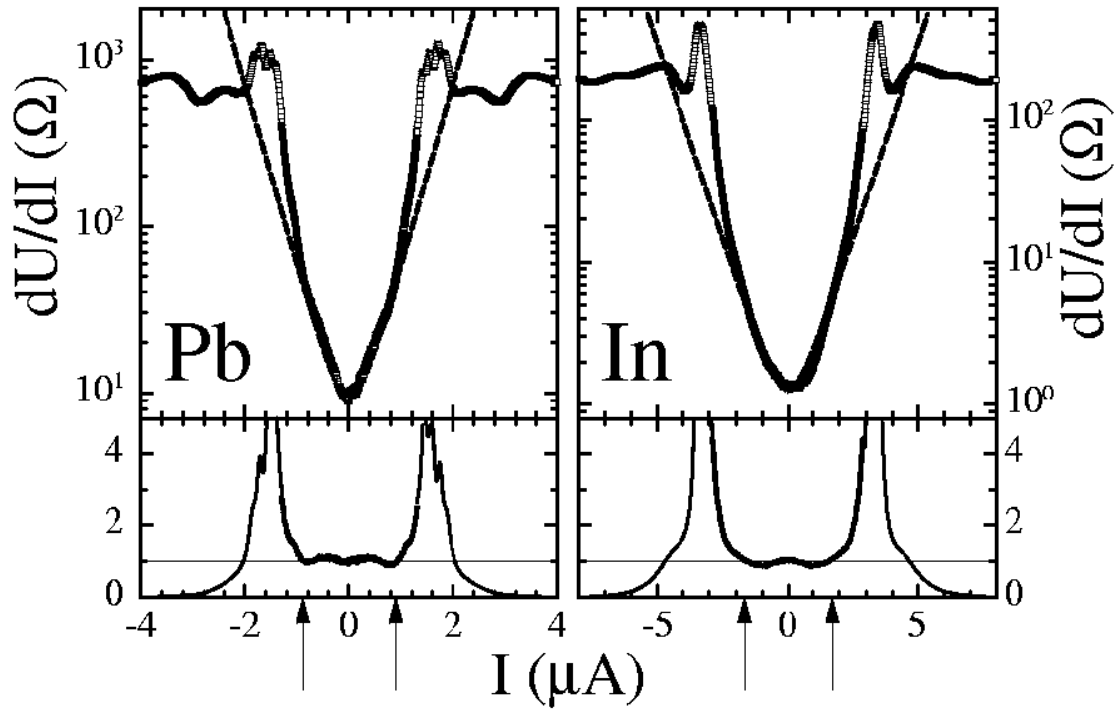,width=150mm}\hfill
\caption[zener-x]{\small\rm
 $dU/dI$ vs.~$I$ at $T=50\,$mK of a Pb and an In junction. The dashed lines 
 are fits using Eq.~\ref{dudi}. 
 The lower two diagrams show the experimental data normalized to the fit.
 Arrows mark the current $I^*$ below 
 $I_{\rm{c}}$ at which the measured curves starts to deviate from the fit.}
\label{zener}
\end{figure}

\subsection{Zener tunneling of the phase}
\begin{figure}
\hspace{20mm}\psfig{file=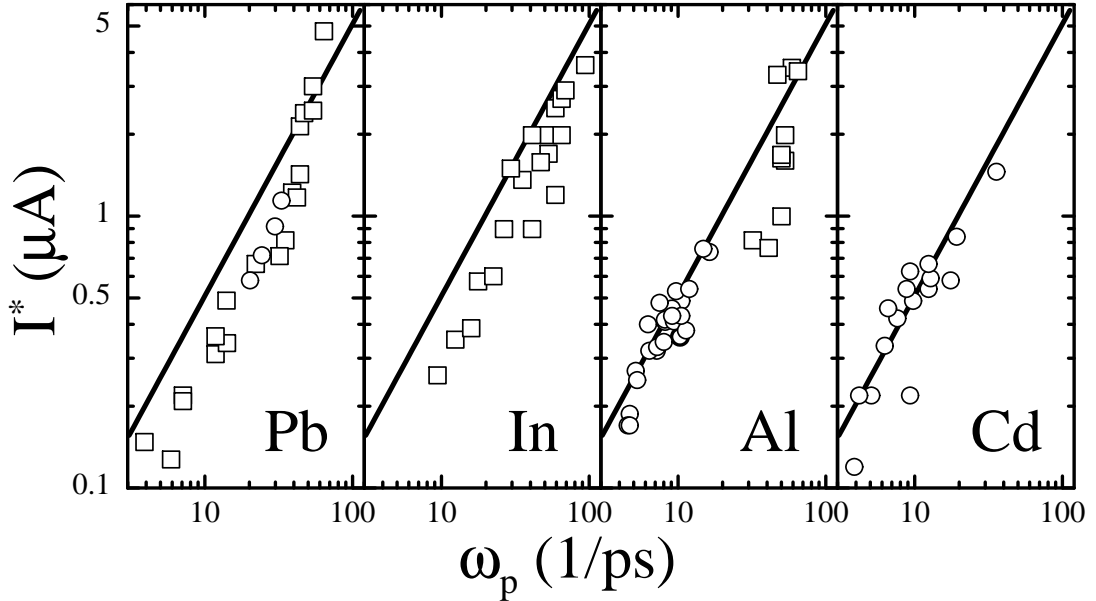,angle=270,width=150mm}\hfill
\caption[zener-freq-x]{\small\rm
 $I^*$ vs.~plasma frequency $\omega_{\rm{p}}$. The solid line is 
 $I^* = I_{\rm{Z}}$. The squares mark the anomalies attributed to Zener 
 tunneling of the phase, circles denote the position of the minima of the
 spectra.} 
 \label{zener-freq}
\end{figure}

$E_{\rm{JE}}$ and $\omega_{\rm{p}}$ have been derived from the spectra at 
rather small currents. The intermediate range 
below the experimental $I_{\rm{c}}$ usually shows a steep rise of $dU/dI$. 
Some of the junctions, especially those with Pb and In, even have a distinct 
kink at $I^*\approx 0.2 I_{\rm{c}}^0$ (Fig.~\ref{zener}). Such a steep rise 
has to be expected, and it is straightforwardly explained as Zener tunneling 
of the phase at $I^*  = I_{\rm{Z}}$, when the lowest energy level of one of 
the potential wells exceeds the next higher level of the neighbouring well. 
This is the more important since $E_{\rm{JE}}\approx \hbar\omega_{\rm{p}}$
of these junctions implies that the  washboard potential contains not more
than two or three discrete levels. Fig.~\ref{zener-freq} shows that the
agreement between $I^*$ and $I_{\rm{Z}}$ is quite good for large
$\omega_{\rm{p}}$ (and small $R_0$). At small frequencies (and large $R_0$),
the current $I^*$ falls below $e\omega_{\rm{p}}/\pi$. This has to be 
expected, too, because the level spacing of the anharmonic washboard 
potential decreases when $E_{\rm{JE}}/\hbar\omega_{\rm{p}}$ decreases.

Nevertheless, Zener tunneling of the phase implies that the junctions are 
only weakly damped. Otherwise, the experimental $R_0$ would yield  
$E_{\rm{JE}}/\hbar\omega_{\rm{p}}\le 0.5$, and the washboard potential 
would have not more than one discrete level. As Zener tunneling features 
could be observed at junctions with $R_0$ up to about 260, 175, and 
$15\,\Omega$ for Pb, In, and Al, respectively, the quality factor $Q\ge 5$.
This supports our interpretation in the previous section.

Zener tunneling of the phase is preferably resolved at Pb or In junctions 
and not at Al or Cd junctions. The simplest explanation is that the voltage 
drop at $I = I_{\rm{Z}}$ 
\begin{equation}
U_{\rm{Z}} \approx \frac{\sinh (3.6)}
   {3.6\pi}\,\, e \omega_{\rm{p}}R_0
\end{equation}
of about $0.26\,\mu{\rm{V}}\cdot\omega_{\rm{p}}(1/{\rm{ps}})R_0(\Omega)$ 
has to be considerably smaller than the superconducting gap $2\Delta/e$. 
This condition is the easier fulfilled the larger the energy gap is, that 
is for Pb and not for Cd.

\begin{figure}
\hspace{20mm}\psfig{file=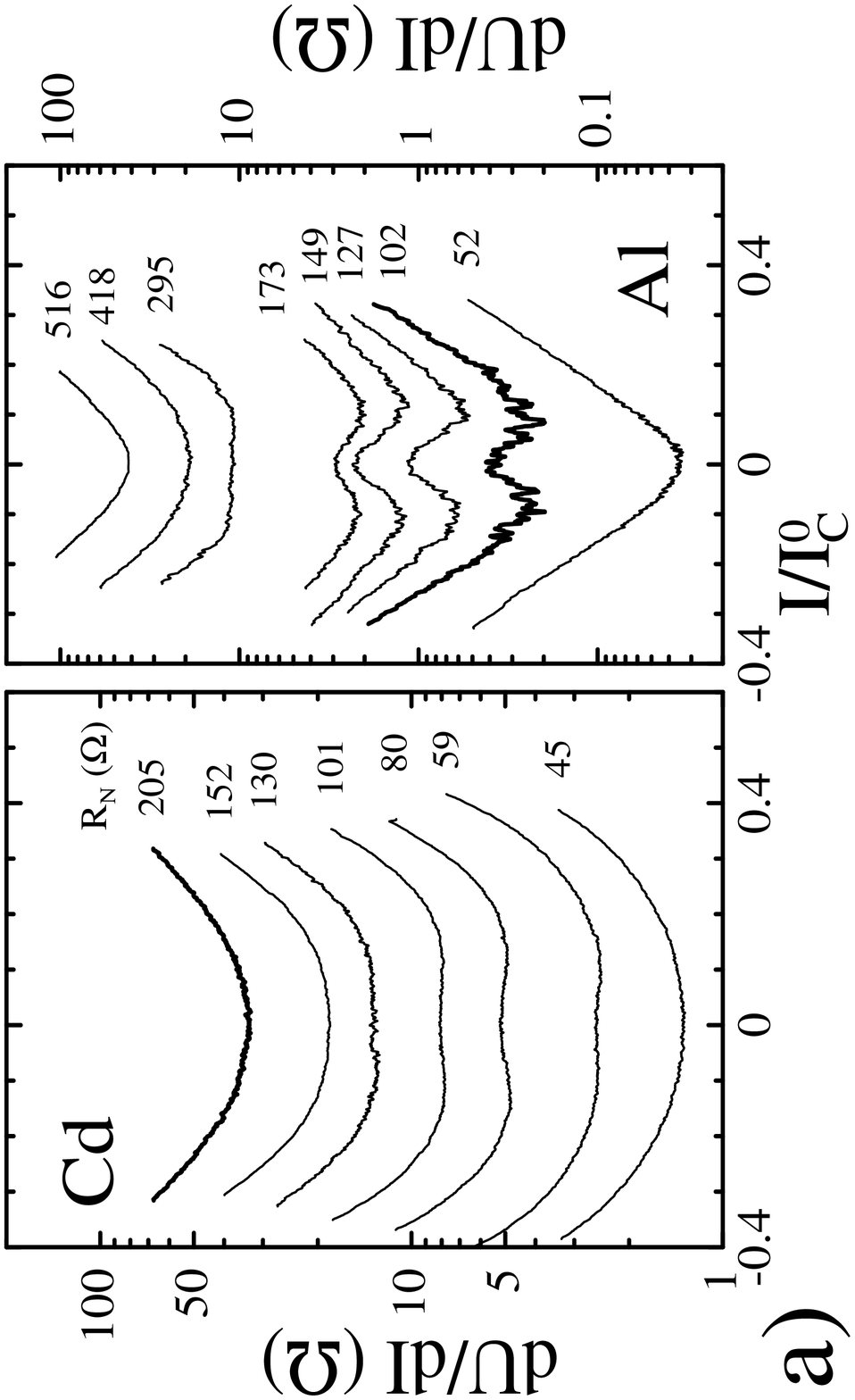,angle=270,width=150mm}\hfill

\hspace{20mm}\psfig{file=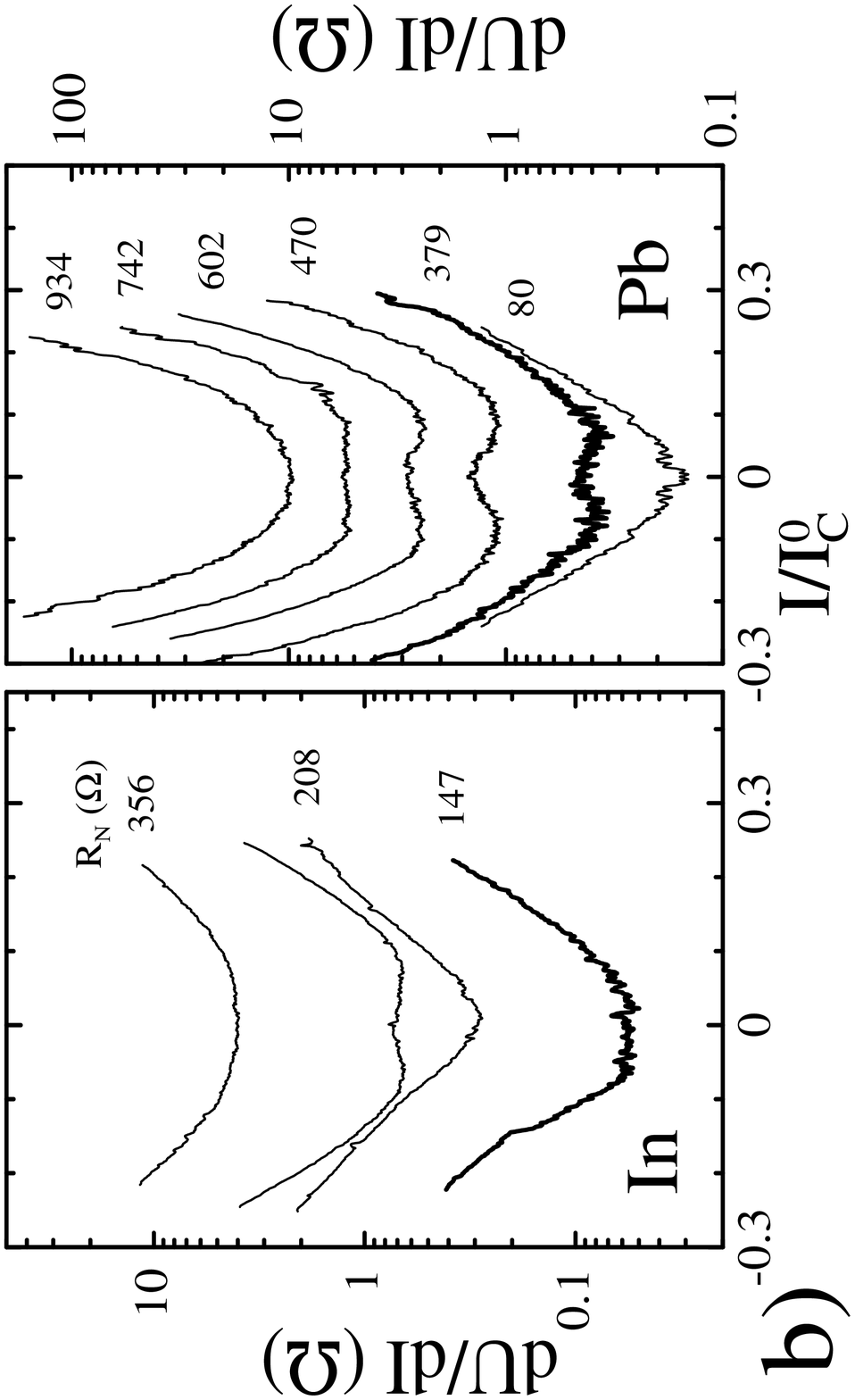,angle=270,width=150mm}\hfill
\caption[bloch-x]{\small\rm
 $dU/dI$ vs.~$I/I_{\rm{c}}^0$ at $T=50\,$mK of some selected junctions
 showing additional double-minimum structures around $I^*=0.1\,I_{\rm{c}}^0$. 
 $I_{\rm{c}}^0$ with the clean-limit critical current.}
\label{bloch}
\end{figure}

\subsection{Possible excitation of Bloch waves} 
One source of uncertainty in deriving both $E_{\rm{JE}}$ and 
$\omega_{\rm{p}}$ are additional structures -- zero-bias maxima -- superposed 
on $dU(I)/dI$ of some of the junctions (Fig.~\ref{bloch}). They
were observed quite often for Al and Cd and less often for Pb and In. 
Conventional explanations for such zero-bias maxima are Kondo scattering 
at magnetic impurities or Andreev reflection at a normal-superconducting 
interface. Both explanations seem to be feasible here, because they only 
require a small normal interfacial region or a tiny amount of Kondo 
impurities (on the ppm level), that may have been introduced by cutting 
the nut. But these two phenomena should take place at nearly constant 
voltage, while the experimental data are clearly related with a current 
of $I^*/I_{\rm{c}}^0 \approx 0.1$.

An alternative explanation is Coulomb blockade with its negative 
contribution to $dU/dI$, and one may speculate whether these additional 
anomalies in Fig.~\ref{bloch} result from the excitation of Bloch waves, 
like in Ref.~\cite{Kuzmin91}. In favour of this interpretation is that
in the previous paragraph we have found hints that the damping is 
smaller than expected and the quasi-particle resistance much larger than 
$R_{\rm{N}}$. However, our junctions have Coulomb charging
energies $E_{\rm{C}} = (\hbar\omega_{\rm{p}})^2/8E_{\rm{JE}} 
\approx E_{\rm{JE}}/10$, that means they are in the tight-binding limit. 
If there was a Coulomb blockade then it can not be expected to be well 
pronounced, like what is seen in Fig.~\ref{bloch}.
The transport of Cooper pairs could also be enhanced when the periodic 
transfer of Cooper pairs is in resonance with the zero-point oscillations 
of the Josephson plasma. The position $I^*$ of the minima 
marks then the Zener current $I_{\rm{Z}}$ for tunneling of the quasi-charge.
As a cross-check, we estimate $E_{\rm{JE}}$ from $R_{\rm{N}}$ by assuming 
the theoretical $I_{\rm{c}}^0$ and calculate then $\omega_{\rm{p}}$ from the 
experimental $R_0$. Indeed, the position $I^*$ of the minima fits reasonably 
well the above estimated plasma frequency, that is  $\pi I^*/ e 
\approx \omega_{\rm{p}}$, see Fig.~\ref{zener-freq}. 

A negative contribution could be superposed on the $dU/dI$ spectra of 
the other junctions, even if a minimum cannot be resolved. The relative
size of such a contribution may depend on damping, that is on the (unknown)
quasi-particle resistance $R_{\rm{qp}}$. The plasma frequency and, 
consequently, the coupling energy would then have been overestimated.
However, in view of the available experimental data, these possible
corrections are difficult to handle.

\subsection{Noise}
Up to now we have neglected electrical noise. Since noise can also drive 
the junctions towards the resistive state, we have to ensure that it is 
small enough. A natural measure for the noise magnitude is the 
superconducting gap $2\Delta$. Low-frequency noise up to about 2 MHz makes 
no problem, we can detect it directly: including amplifier noise,
it is less than about $2\,\mu$eV ($10\,\mu$eV) at a $100\,\Omega$ 
($10\,{\rm{k}}\Omega$) resistor and comparable to the thermal noise 
$k_{\rm{B}}T \approx 4\,\mu{\rm{eV}}$ at our lowest temperature of 50 mK. 
This is much smaller than $2\Delta$ of the four superconductors, and
it is consistent with the fine structure of the spectra due to multiple
Andreev reflection in Fig.~\ref{spectra}.

The possible radio-frequency component of the external noise, however, can 
only be revealed indirectly through the properties of the junctions. For 
this reason we can not {\em prove} that this sort of noise is negligible.
Let us assume, for the moment, that the finite $R_0$ as well as the reduced
$I_{\rm{c}}$  mainly result from external noise. What is then the minimum
noise level required to explain these results ? Weakly damped
junctions in the thermally activated regime are used for this estimate
because including damping and assuming quantum tunnelling would
require more noise. The tunneling rate is then given by the well-known
Arrhenius law 
\begin{equation}
\Gamma_{\rm{TA}} = \frac{\omega_{\rm{p}}}{2\pi}
  \exp{\left( -\frac {2E_{\rm{JE}}} {k_{\rm{B}}T_{\rm{eff}}} \right)}
\label{ta-rate}
\end{equation}
$T_{\rm{eff}}$ is an effective temperature $T_{\rm{eff}}=T+T_{\rm{noise}}$,
that is we assume white noise with an equivalent temperature
$T_{\rm{noise}}$, and superpose it on the thermal noise of the junctions.
Using the same arguments as for quantum tunneling
Eqs.~\ref{qtrate}-\ref{derivative}, the differential resistance at
$I\rightarrow 0$ 
\begin{equation}
 dU(I)/dI \approx
   R_0 \cosh{\left(\frac{\pi\hbar}{2e k_{\rm{B}}T_{\rm{eff}}} I \right)}
\label{ta-dudi}
\end{equation}
Here the residual resistance $R_0 = dU/dI(I=0)$ 
\begin{equation}
 R_0 \approx 
   R_{\rm{K}} \frac {\hbar \omega_{\rm{p}}} {4 k_{\rm{B}}T_{\rm{eff}}}
   \exp{\left(-\frac{2 E_{\rm{JE}}}{k_{\rm{B}}T_{\rm{eff}}} \right)}
\label{ta-residual}
\end{equation}
The second-order approximation of Eq.~\ref{ta-dudi}
\begin{equation}
\frac{dU(I)}{R_0 dI} \approx 1 + \frac{1}{2} \left(\frac{\pi\hbar I}
  {2 e k_{\rm{B}}T_{\rm{eff}}}\right)^2 
\label{ta-derivative}
\end{equation}
now allows to derive the effective temperature of the junctions from the
spectra at $I\rightarrow 0$. By comparing the $I^2$-dependence of
Eq.~\ref{ta-derivative} with that of Eq.~\ref{derivative}, we can
directly read off
\begin{equation}
k_{\rm{B}} T_{\rm{eff}} = \hbar \omega_{\rm{p}}/3.6 
\label{noise}
\end{equation}
from the data in Fig.~\ref{analysis-wp} (the $\omega_{\rm{p}}$ in
Eqs.~\ref{ta-rate} and \ref{noise} denote different parameters).
This yields {\em lower} bounds for
the hypothetical noise temperature in the 1 - 100 K range. If such a noise
level was real, we would never have seen any superconducting features of
our samples. Thus we conclude that the external noise level is much
smaller than the zero-point energy of the junctions, and $\omega_{\rm{p}}$
in Fig.~\ref{analysis-wp} is the plasma frequency.

The shape of the $I(U)$-characteristics at $I_{\rm{c}}$ responds more
sensitively to external noise than the residual resistance. The pronounced
kink at the critical current of an ideal $I(U)$-characteristic can be
rounded, and both the hysteresis and $I_{\rm{c}}$ can be reduced when the
noise level exceeds even a small fraction of the coupling energy $E_{\rm{JE}}$,
see for example Ref.~\cite{Ambegaokar69}.
To observe this noise, its size must be at least comparable to the
zero-point energy of the particle in the tilted washboard potential. Since
increasing the current means to reduce the minimum height of the potential
wells $2E$ as well as the zero-point energy $\hbar\omega/2$, at some current
below the intrinsic critical one, thermal or external noise should
definitely overtake. However, in contrast to the analytical Eq.~\ref{plasma},
the zero-point energy has a lower bound as described by Eqs.~\ref{eje-wp}
and \ref{reduced-ic}. This is due to the fact that the flow of supercurrent 
needs at least one discrete energy level.
Those junctions with finite $R_0$, for which we can derive the plasma
frequency, would require noise temperatures in the 1 - 1000 K range. Such a
noise level is by far too large, as found above while discussing the
possible effects of noise on $R_0$ and the $I(U)$-characteristic at
$I\rightarrow 0$. Towards smaller contact resistances, the experimental
$E_{\rm{JE}}/\hbar\omega_{\rm{p}}$ ratio increases slightly, the critical
current approaches the theoretical $I_{\rm{c}}^0$ (at least for Pb and In),
and the minimum $\omega$ can be much smaller than $\omega_{\rm{p}}$.
But the absolute value of  $\omega_{\rm{p}}$ also rises if we extrapolate
the data in Fig.~\ref{analysis-wp}, so the necessary noise level will not
go down.
The hysteresis found at some of the low-resistance junctions in the
$R_{\rm{N}} = 1 - 10\,\Omega$
range could be explained by low damping, although heating effects can not be
exluded. At larger contact resistances, the minimum required noise level
becomes smaller than about $T_{\rm{noise}}\approx 1\,$K. It is then quite
possible that external noise contributes to the spectra.
We conclude that the rounding of the $I(U)$-characteristics
as well as the reduction of $I_{\rm{c}}$ mainly origines from the
quantum-mechanical zero-point fluctuations as an {\em{internal}} noise
source of our Josephson junctions and not from external noise. Of course,
it is this internal noise that can also strongly reduce the hysteresis of
the $I(U)$-characteristics.

There are additional arguments that noise can indeed be neglected:

a) Our experimental observations agree qualitatively with previous ones 
by others who used a similar type of setup for preparing the break junctions 
\cite{Muller92,Muller94,Peters95,vanderPost97}. If external noise was the
limiting factor, then the noise level should be the same in both experiments.
But it is unlikely to have an identical noise level at these various
experiments and at different times and locations.

b) A different series of break-junction experiments on doped semiconductors 
(Germanium) using our setup indicated a noise level of less then about
$50\,\mu$V for junctions with $R_{\rm{N}} < 100\,{\rm{k}}\Omega$, and
corresponding to a noise temperature of less than about 0.5 K. Thus the
total noise at the junctions consist mainly of the low-frequency component.

c) If the high-frequency noise would dominate, it would roughly be 
related to the experimental zero-point energy $\hbar\omega_{\rm{p}}/2$,
as discussed above. We expect junctions with larger $R_{\rm{N}}$ to pick
up more noise then those with smaller  $R_{\rm{N}}$, independent of the
superconducting gap. Quite contrary, the experimental $\omega_{\rm{p}}$
in Fig.~\ref{analysis-wp} {\em decreases} almost like $\Delta/R_{\rm{N}}$.
Note that for these junctions $\hbar\omega_{\rm{p}}/2 \ge 2\Delta$.

d) Both the experimental critical current $I_{\rm{c}}$ from $I(U)$ at 
large currents and the $I(U)$ characteristic at $I \rightarrow 0$ 
consistently indicate a small capacitance of order 0.1 fF.

e) The experimental $E_{\rm{JE}}$ derived from $dU(I)/dI$ at 
$I \rightarrow 0$ nearly coincides with the theoretical coupling energy
(Fig.~\ref{analysis-eje}). There is no adjustable parameter.

f) Although rather different processes (Zener tunneling of the phase as well
as the excitation of Bloch waves and Zener tunneling of the quasi-charge)
may contribute to the spectra at $I^* \approx I_{\rm{Z}}$, this marker
closely corresponding to $e\omega_{\rm{p}}/\pi$ supports our derivation of
the plasma frequency. 

We believe these arguments as a whole strongly support the basic 
concepts of our interpretation. What is needed now is to understand the
physical meaning of the capacitance of metallic Josephson junctions.

\subsection{The horizon model of tunnel junctions}
Future research on superconductors with even lower $T_{\rm{c}}$, those that 
have gaps with nodes of zeroes, or gaps with unknown symmetry like the
heavy-fermion superconductors, requires to reduce the quantum fluctuations 
by increasing the capacitance. 

It is difficult to extract the capacitance of solitary junctions. This 
problem has been discussed extensively for the case of Coulomb blockade at
tunnel junctions, see for example
Refs.~\cite{Delsing89,Flensberg91,Wahlgren95,Hirvi97}, applying Nazarov's
horizon model \cite{Nazarov89}.
Our break junctions in the vacuum-tunneling regime, when the two 
halves of the sample are not in direct contact but are separated by a 
$0.1 - 1\,$nm wide vacuum gap, and in the normal state have an asymptotic 
offset of their $I(U)$ characteristic at $\sim 50\,$mV in the range 
0.2 - 2 mV. Attributing this offset to Coulomb blockade, we estimate  
an intrinsic capacitance $C_0$ of order 0.1 fF due to the vacuum gap.

\begin{figure}
\hspace{20mm}\psfig{file=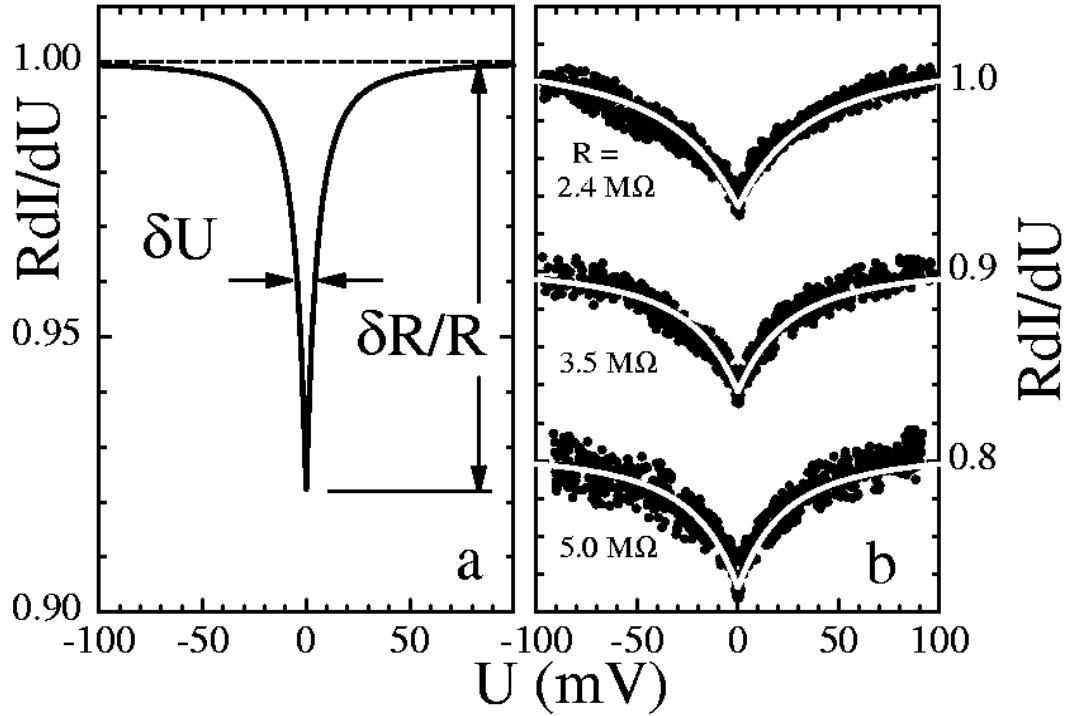,width=150mm}\hfill
\caption[tunnel-spec-x]{\small\rm
 a) Normalized conductance of a tunnel junction with a constant 
 $1/R=dI/dU$ (dashed line), modified by Coulomb blockade Eq.~\ref{coulomb} 
 (solid line). The relative size $\delta R/R$ of the zero-bias anomaly 
 determines the lead capacitance $\kappa$, assumed to be $\kappa = 5\,$pF/m.
 The capacitance $C_0$, here 0.1 fF, is inversely proportional to both the 
 width $\delta U$ and $\delta R /R$.
 b) Experimental data of Al junctions in the normal state at $T=50\,$mK. 
 Superconductivity has been suppressed by a magnetic field of $B=100\,$mT. 
 $R$ is the differential resistance at $U \approx 100\,$mV. The two lower
 spectra have been shifted for clearity. The solid lines are fits using 
 Eq.~\ref{coulomb} with $\kappa \approx 5 - 6\,$pF/m. The intrinsic 
 capacitance $C_0 \approx 20\,$aF.}
\label{tunnel-spec}
\end{figure}

At smaller voltages the lead capacitance enhances the total capacitance with 
respect to $C_0$. The leads -- which means here not only the current and 
voltage leads but also the two halves of the sample and the close 
surroundings of the junction itself -- can be treated as a transmission line 
with a capacitance per length $\kappa$. The Heisenberg uncertainty principle 
now defines the shortest time scale for the relevant processes at the 
junction, and thus a maximum spread or horizon of about \cite{Nazarov89}
$c\hbar/e|U|$ (at $U=1\,$mV this spread amounts to about 188 $\mu$m). Here
$c$ denotes the speed of light. The total capacitance becomes then
\begin{equation}
C(U) \approx C_0 + \frac{\kappa c \hbar}{e|U|}
\label{capacity}
\end{equation}
diverging at $U \rightarrow 0$. 
To incorporate Coulomb blockade into the low-temperature spectrum of a
tunnel junction with constant conductance $1/R = dI(U)/dU$, its $I(U)$ 
characteristic is displaced by $e/2C(U)$ on the voltage axis, that is
\begin{equation}
R I \approx U -\frac{e}{2C(U)} \, {\rm{sign}}(U)
\end{equation}
The normalized conductance becomes
\begin{equation}
R \frac{dI}{dU} \approx 1 - \frac{e^2\kappa c \hbar} 
           {2\left( e |U| C_0 + \kappa c \hbar \right)^2}
\label{coulomb}
\end{equation}
At large voltages $dI/dU = 1/R$ is recovered, but there is a $\delta R/R =
e^2/2\kappa c \hbar$ deep mimimum at $U=0$, see Fig.~\ref{tunnel-spec}. 
With this dip the lead capacitance amounts to
\begin{equation}
\kappa  \approx \frac{e^2}{2 c \hbar}\, \frac{R}{\delta R}
\label{kappa}
\end{equation}
$C_0$ can then be estimated from the width $\delta U$ of the anomalies. 
For our vacuum tunnel junctions (Fig.~\ref{coulomb}) we get an average 
$\kappa \approx 5\,$pF/m (for the different junctions $\kappa$ was found to 
vary between about 2 pF/m and 10 pF/m). This is a quite reasonable value, 
although being smaller than typical literature data reported for tunnel
junctions, for example $\kappa\approx 10\,$ pf/m \cite{Delsing89} and
$\kappa \approx 20 - 32\,$pF/m \cite{Wahlgren95}. 

\begin{figure}
\hspace{20mm}\psfig{file=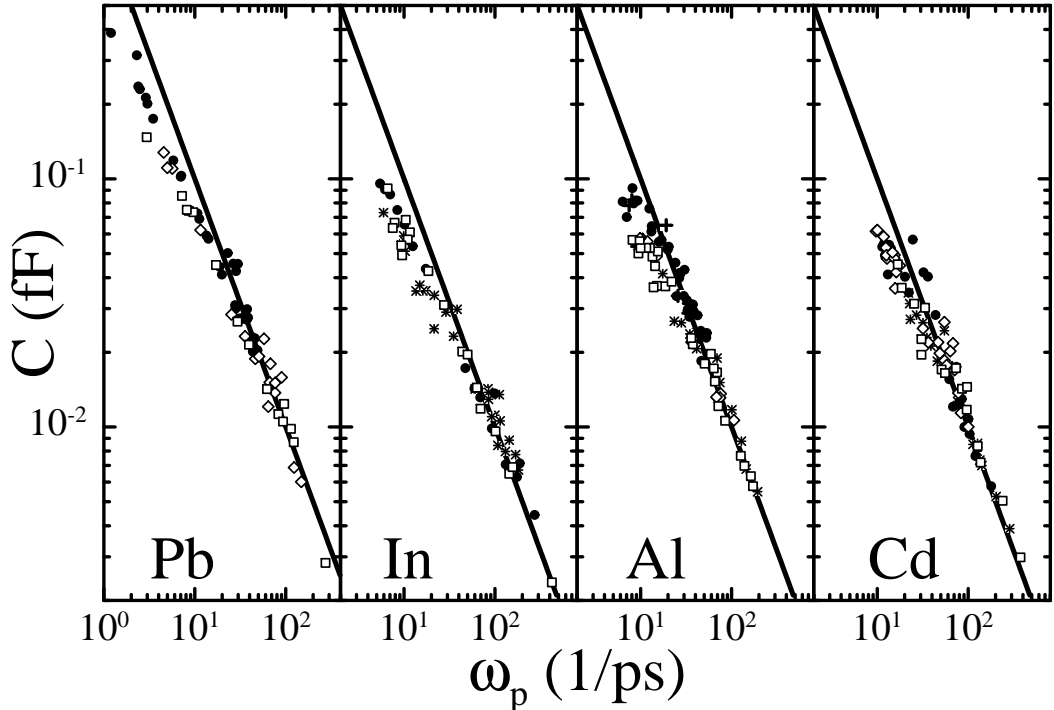,angle=270,width=150mm}\hfill
\caption[c-of-wp-x]{\small\rm
 Capacitance $C=4e^2E_{\rm{JE}}/\hbar^2\omega_{\rm{p}}^2$ vs.~plasma 
 frequency $\omega_{\rm{p}}$ of the Josephson junctions. The solid lines 
 represent $C=\kappa c / \omega_{\rm{p}}$ with $\kappa = 3.3\,$pF/m.}
\label{c-of-wp}
\end{figure}

\subsection{The dynamic capacitance of metallic Josephson junctions}
Metallic junctions have an intrinsic $C_0 = 0$, that is only the dynamic 
capacitance contributes. The small experimental capacitances in
Fig.~\ref{analysis-c} together with the lead capacitance $\kappa$ estimated
in the previous section reveal an horizon of $C/\kappa \approx
1-100\,\mu$m, restricted to the immediate vicinity of the junctions. The
huge capacitances ($\sim 0.1\,$ nF) of the external current and voltage
leads do not matter, in contrast to the interpretation in
Refs.~\cite{Peters95,vanderPost97}. How to derive theoretically the horizon
of a ballistic Josephson junction ?

The theory for tunnel junctions \cite{Nazarov89,Flensberg91} is based
on the idea that the charge transfer across the junction creates an
excess charge on the other lead. This excess charge is transported down
the leads via photons with arbitrary low energy. The new state is
orthogonal to the equilibrium ground state, yielding an infrared
divergency. In the quantum regime of the junction, the divergency
manifests itself as a zero-bias anomaly with a power law.
We do not know whether such a theory can be applied to metallic Josephon
junctions investigated here. However, we will use the physical idea that
the infrared divergency, caused by the interaction of the system with its
electromagnetic environment, is cut off at the lowest characteristic
frequeny of the problem.

The RCSJ model maps the many degrees of freedom of the electron system
involved in  charge transfer across a Josephson junction onto a single
degree of freedom, the phase difference $\varphi$ between the two bulk
superconductors. Since phase and charge density are canonically conjugated
variables, and the electromagnetic gauge field couples to the charge, its
quantum fluctuations modify dynamically the capacitance of the junction.
Moreover, the interactions with the electromagnetic field are strong enough
to {\em define} the capacitance (or mass) of the particle in the washboard
potential. The ground state of the quantum-mechanical model of a Josephson
junction even at finite current $I < I_{\rm{c}}$ is a meta-stable one. It
can be described by the ground state of an harmonic oscillator with the
Josephson plasma frequency $\omega_{\rm{p}}$ and a zero-point energy
$\hbar \omega_{\rm{p}}/2$. A constant expectation value $<\varphi>$ leads
to $U = 0$, although the charge fluctuations of the ground
state take place on a time scale $1/\omega_{\rm{p}}$. Since this is the only
time scale of the system, we expect $\omega_{\rm{p}}$ to play the crucial
role as cut-off frequency in defining the horizon. And this will yield a
self-consistent picture for the interpretation of our experimental results.

A different approach is known in quantum mechanics: a conventional particle
of mass $m$ interacting with a relativistic quantum field. This interaction
includes a certain environment around the particle, limited by a cut-off
parameter $\sim c \hbar/m c^2$ \cite{Messiah79}. This cut-off --  which is
just the horizon we are looking for -- depends on the internal energy
$mc^2$ of the particle. It serves to avoid the creation of additional
particles, excluded in the non-relativistic treatment of the particle itself.
Applying this model to a Josephson junction lacking a real mass, the
internal energy $mc^2$ has to be replaced by the excitation energy
$\hbar\omega_{\rm{p}}$ to avoid transitions between the ground state and
the next higher level.

Both models thus predict the same horizon $\sim c/\omega_{\rm{p}}$. The total
capacitance becomes then
\begin{equation}
C \approx \kappa c / \omega_{\rm{p}}(C)
\label{horizon}
\end{equation}
Note that $\omega_{\rm{p}}$ is a function of $C$ with a {\em positive} 
feedback: the higher the frequency the smaller is the  capacitance, 
further enhancing the plasma frequency. Solving for $C$ and 
$\omega_{\rm{p}}$ yields
\begin{equation}
 C \approx \frac{\kappa^2 c^2 \hbar}{2 e I_{\rm{c}}^0}
\label{theo-cap}
\end{equation}
and
\begin{equation}
\omega_{\rm{p}} \approx \frac{2 e I_{\rm{c}}^0}{\kappa c \hbar}
\label{wp-ic}
\end{equation}
respectively. The experimental data in Figs.~\ref{analysis-wp} and
\ref{analysis-c} reproduce indeed such a dependence from $I_{\rm{c}}^0
\propto 1/R_{\rm{N}}$. The absolute values of $C(\omega_{\rm{p}})$ of 
the four investigated superconductors in Fig.~\ref{c-of-wp} fit almost 
perfectly a $\kappa \approx 3.3\,$pF/m. The good agreement with $\kappa
\approx 5\,$pF/m estimated above for the tunnel junctions, supports
our interpretation.

Since frequency and capacitance are related to each other, 
Eqs.~\ref{theo-cap} and \ref{wp-ic} are valid only at $I\rightarrow 0$, like 
Eq.~\ref{exp-cap}. When the applied current approaches the intrinsic 
$I_{\rm{c}}^0$, the capacitance should become larger because the plasma 
frequency decreases. Near $I_{\rm{c}}^0$ the capacitance should diverge 
(this is partly compensated by the applied current introducing an additional 
time scale or frequency $\sim I/$e). For this reason the relation 
Eq.~\ref{icrn} can not adequately describe the experimental data (disregard 
the correct order of magnitude of $C$), that is quantum fluctuations only 
indirectly suppress the critical current. 

Most of the critical current data in Fig.~\ref{rnic} have been reduced with 
respect to the theoretical $I_{\rm{c}}^0$. Therefore it seems reasonable
to assume $C(I_{\rm{c}}) \approx C(I \rightarrow 0)$. Phase diffusion at 
$|\dot{\varphi}| \ge \omega_{\rm{p}}$ becomes then responsible for 
$I_{\rm{c}}$ being too small. As the Zener current 
$I_{\rm{Z}}=e\omega_{\rm{p}}/\pi=4I_{\rm{c}}^0/\kappa c R_{\rm{K}}$, the 
critical current at $Q\gg 1$ amounts to
\begin{equation}
  I_{\rm{c}}  \approx    
   \frac {I_{\rm{c}}^0} {0.9 \kappa c R_{\rm{K}}  }\,
   {\rm{arcsinh}} \left( \frac{1.8 R_{\rm{K}}}{{R_0}} \right)
\label{ic-of-r0-kappa}
\end{equation}
This $I_{\rm{c}}$ depends directly on the residual resistance. 
Eq.~\ref{ic-of-r0-kappa} fits the experimental data quite well, see Fig. 
\ref{ic-r0}. No adjustable parameter is required because the lead
capacitance $\kappa$ is known from the spectra at $I\rightarrow 0$.
Considering damping reduces the critical current of Eq.~\ref{ic-of-r0-kappa} 
by a factor of about $\left[1+0.87/Q\right]$. The close coincidence 
between $R_{\rm{N}}I_{\rm{c}}(R_0)$ and Eq.~\ref{ic-of-r0-kappa} thus 
excludes any strong effect from damping. It supports our above estimate of
$Q=\omega_{\rm{p}}R_{\rm{qp}}C = \kappa c R_{\rm{qp}} \ge 1$, and the 
quasi-particle resistance of the ballistic junctions being considerably 
larger than $R_{\rm{N}}$.

There seem to be several possibilities to reduce the lead capacitance 
$\kappa$ and drive the junctions towards the delocalized Coulomb blockade 
limit, or to make it larger and get more localized
Josephson-like behaviour. To enlarge the lead capacitance $\kappa$ by 
increasing the accessible volume or by bringing the junctions closer to 
the conducting ground plane than is possible with bulk samples, one should 
use junctions made of thin films like that described in Ref.~\cite{Scheer97} 
or whiskers. On the other hand, 
the conduction electrons around the junction of a bulk sample reduce the 
effective volume of the capacitor, as long as $\omega_{\rm{p}}$ is smaller 
than the plasma frequency $\sim 10^4\,{\rm{ps}}^{-1}$ of the conduction 
electron system, as already proposed by K.~K.~Likharev \cite{Likharev79}. 
In an extreme limit, the Josephson junction is completely embedded in normal 
metal. Alternating currents and fluctuations could then exist only at 
frequencies above the plasma frequency of the conduction-electron system, 
enhancing the quantum fluctuations by driving the effective capacitance 
towards zero. Such junctions with strongly reduced critical currents may 
have been realized already with the so-called heavy-fermion superconductors. 
At junctions with these superconductors a considerable part of the contact 
region seems to be driven normal due to stress and disorder, and the product 
$R_{\rm{N}}I_{\rm{c}}$ was found to be several orders of magnitude smaller 
than expected, see e.~g.~Ref.~\cite{Gloos-latsize}. 
To take into account a small $C$ may be an interesting 
new aspect for the future study on those superconductors.

\section{Summary}

The quantum-mechanical treatment of the RCSJ model explains quite well the 
properties of superconducting point contacts over a wide range of 
resistances. The small capacitance of the junctions reproduces both the 
suppression of the critical current and the finite contact resistance by 
phase diffusion and drift due to quantum tunneling at the presence of large 
quantum-mechanical fluctuations of the Josephson plasma. Our interpretation 
does not require external noise and the huge capacitances of the current 
and voltage leads. A detailed analysis of the $I(U)$ characteristics and 
$dU(I)/dI$ spectra reveals a frequency-dependent capacitance of the 
junctions that is well described by the horizon model, that has first been
applied to tunnel junctions. The lead capacitance is the only free parameter 
with which one can almost fully understand our Josephson junctions. 
Combining the properties of the metallic Josephson junctions with 
that of a completely different type of vacuum tunneling junctions in the
normal state strongly supports our interpretation. There are several hints 
that the metallic Josephson junctions possess a rather large
quasi-particle resistance $R_{\rm{qp}} \ge 1\,{\rm{k}}\Omega$. They
are therefore only weakly damped, and anomalies which we have 
attributed tentatively to the excitation of Bloch waves, Zener tunneling of 
the quasi-charge, and Zener tunneling of the phase become plausible
even though the normal-state resistance of the junctions is less than
$R_{\rm{K}}/4$.

\section*{Acknowledgments}
F.~A. thanks A.~van Otterlo for discussions.
This work was supported by the SFB 252 Darmstadt/Frankfurt/Mainz
and the German BMBF Grant No. 13 N 6608/1.


\end{document}